\documentclass[multphys,vecphys]{svmult}
\usepackage{makeidx}     
\usepackage{graphicx}    
\usepackage{multicol}    

\def\be{\begin{equation}}
\def\ee{\end {equation}}
\def\bea{\begin{eqnarray}}
\def\eea{\end{eqnarray}}
\def\bml{\begin{mathletters}}
\def\eml{\end{mathletters}}
\def\l{\label}

\def\ra{\rightarrow} 

\begin{document}

\title*{Adaptation in simple and \\ complex fitness landscapes\thanks{To appear
in \textit{Structural approaches to sequence evolution: Molecules, networks
and populations}, ed. by U. Bastolla, M. Porto, H.E. Roman and M.
Vendruscolo (Springer, Berlin 2006).}}
\author{Kavita Jain\inst{1} \and Joachim Krug\inst{1,2}}
\institute{1. Institut f\"ur Theoretische Physik, Universit\"at zu
K\"oln, Germany \\ 
2. Laboratory of Physics, Helsinki University of Technology,
Finland \\
\texttt{kavita@thp.uni-koeln.de, krug@thp.uni-koeln.de}}

\maketitle


\section{Introduction}

The notion that evolution can be viewed as a hill-climbing
process in an adaptive landscape was introduced in 1932 by 
Sewall Wright \cite{Wright32}, and remains one of the most powerful 
images in evolutionary biology \cite{Gavrilets04}. 
Since the discovery of the molecular
structure of genes it has been clear that the substrate over which
the adaptive landscape should be properly defined is the space of genetic
sequences \cite{Smith70}. Nevertheless, apart from a few landmark
papers \cite{Eigen71,Kauffman87}, adaptation has not been in the focus
of the theory of molecular evolution, which instead has concentrated on 
the effects of stochastic drift in a neutral (\textit{flat}) fitness
landscape \cite{Crow70}. This situation is presently changing
\cite{Travisano01,Orr05}. Long-term evolution 
experiments on microbial populations \cite{Elena03}
are beginning to produce a wealth of data,
on the phenotypic as well as on the genotypic level, which make it
meaningful to ask precise questions about the timing and size of 
adaptive events, and what they can tell us about the structure of 
the underlying adaptive landscape. 

In this chapter we introduce a class of sequence-based models of adaptation, 
which
have been the subject of much recent interest in theoretical population 
genetics
as well as in biologically inspired statistical physics. These models describe
the behavior of a population of haploid, asexual
individuals, each characterized by a genetic sequence
of fixed length, in an adaptive landscape which assigns a fitness value
to each genotype. The population is exposed to the competing
influences of \textit{mutations}, which tend to increase the genetic 
variability,
and \textit{selection}, which focuses the population in regions of high 
fitness. 
The dynamics is deterministic, which implies that the genetic drift induced
by the stochastic sampling noise in finite populations is neglected, and the
adaptive landscape is generally taken to be time-independent. In view of the 
vastness of the field, the selection of topics is unavoidably biased by the 
interests
and preferences of the authors. For a more comprehensive coverage we refer 
the reader
to several recent review articles 
\cite{Baake00,Baake01,Drossel01,Peliti96,Peliti97}. 

The chapter is organised as follows. In the next section the key concepts and 
their
mathematical representation are introduced, and several types of 
mutation-selection
dynamics are described, leaving the form of the adaptive landscape 
unspecified. 
In Sect.~\ref{Simple} we consider simple fitness landscape comprising a single 
adaptive peak or possibly two competing peaks. Here the central theme is the 
\textit{error threshold} phenomenon, which refers to the sudden delocalization
of the population from the fitness peak as the mutation rate increases beyond
a critical value. 
As is described in this book in the chapter by Ester L\'azaro, the
error threshold and the related concept of a \textit{quasispecies} play
an important role in the population dynamics of RNA viruses and in the
development of antiviral strategies.
Due to its similarity to a phase transition, the error 
threshold has been thoroughly analyzed using a range of methods from 
statistical
physics. We give an elementary derivation of the critical mutation rate, and 
describe several
modifications of the basic model, including fitness peaks with a variable 
amount
of epistasis, diploid populations, semiconservative replication, and 
time-dependent
landscapes. 

Section \ref{Complex} is devoted to complex fitness landscapes
consisting of many peaks and valleys. Such landscapes can be modeled by 
ensembles of 
random functions, which links this subject to the statistical physics of 
disordered systems.
Whereas so far the discussion has been restricted to static or steady state 
properties,
time-dependent aspects of mutation-selection dynamics are discussed in 
Sect.~\ref{Dynamics}.
Finally, experimental realizations (\textit{in vitro} as well 
as \textit{in vivo}) of the
models are described in Sect.~\ref{Experiments}, and some concluding remarks 
are presented in Sect.~\ref{Outlook}. 

\section{Basic concepts and models}

In the following discussion, the constituents of a population carry 
a string $\sigma \equiv \{\sigma_{1},...,\sigma_{N}\}$ where each of the $N$ 
letters $\sigma_{i}$ is taken from an alphabet of size $\ell \ge 2$. 
In classical population genetics, 
$\sigma$ represents the configuration of alleles (variants of a gene) 
$\sigma_{i}$ located at gene loci $i$. Typically, 
one-locus, $\ell$ allele models where $\ell$ can take values between two 
(wild type and mutant) to infinity (continuum of alleles) have been considered 
\cite{Burger98}. In the language of population genetics, we are here concerned
with multilocus models with complete linkage \cite{Baake01}.  

At the molecular level, $\sigma$ represents the genetic sequence of an 
individual. For DNA(RNA) based organisms, $\ell=4$ corresponding to 
the nucleotide bases A, T(U), C and G and the sequence length $N$ varies from 
a 
few thousands for viruses to about 
$10^{9}$ for humans. Thus, the total number $4^{N}$ of sequences available  
is hyperastronomically large. The 
minimum value of $\ell=2$  can be obtained by lumping A and G together in 
purins and C, T and U in pyrimidines. The sequences may also 
represent proteins composed of a few hundred amino acids taken from an 
alphabet of size $\ell=20$ \cite{Smith70}.

\subsection{Fitness, mutations, and sequence space}
\l{basic}

The essence of natural selection is that the relative reproductive success of 
an individual
determines whether the corresponding genotype becomes more or less abundant 
in the population.
The \textit{fitness} of an individual is a quantitative measure of its 
reproductive success; depending on the context, it may be defined as
the \textit{viability} of an organism, i.e. the probability to survive to the 
age of reproduction \cite{Gavrilets04}, 
the replication rate of a microbe, the binding affinity of regulatory 
proteins to DNA \cite{Gerland02}
or of antibodies produced by B-cells to pathogens \cite{Perelson95}, 
the program execution speed for digital 
organisms \cite{Wilke02c}, 
or the cost function in an optimization problem \cite{Stadler02}. 

In principle, one should assign fitness to the phenotype which then should be 
related to the genotype; unfortunately, the genotype-phenotype map  
is complicated and largely unknown except for a few cases \cite{Schuster02}
(see also Sect.~\ref{RNAseq}).
This problem is usually outflanked by associating fitness 
$W(\sigma)$ with the genotype itself and define it to be \textit{the expected 
number of offspring produced by an individual with sequence $\sigma$} 
\cite{Peliti97}. 
This definition applies to the case of  discrete generations, 
and is known as \textit{Wrightian} fitness. 
To pass to continuous time dynamics we write
\be
\l{Malthus}
W(\sigma) = \exp[ w(\sigma) \Delta t] \approx 1 + w(\sigma) \Delta t, \;\;\;
\Delta t \to 0,
\ee
where $\Delta t$ is the generation time and 
$w(\sigma)$ is referred to as the 
\textit{Malthusian} fitness \cite{Baake00}.
For future reference we note that multiplication of the Wrightian fitnesses
by a common factor implies a constant additive shift of the Malthusian 
fitnesses.

\begin{table}
\begin{tabular}{cccc}
\hline
\hline
Organism  & Genome size $ \ $ & Rate per base $ \ $  & Rate per genome \\
\hline
Bacteriophage $Q\beta$ & 4.5 $\times 10^{3}$ & 1.4 $\times 10^{-3}$ & 6.5 \\
Vesicular Stomatitis virus & - & - & 3.5 \\
Bacteriophage $\lambda$ & 4.9 $\times 10^{4}$ & 7.7 $\times 10^{-8}$ &
0.0038 \\
\emph{E. Coli} & 4.6 $\times 10^{6}$ & 5.4 $\times 10^{-10}$ & 0.0025 \\
\emph{C. Elegans} & 8.0 $\times 10^{7}$ & 2.3 $\times 10^{-10}$ & 0.018 \\
Mouse  & 2.7 $\times 10^{9}$ & 1.8 $\times 10^{-10}$ & 0.49 \\
Human & 3.2 $\times 10^{9}$ & 5.0 $\times 10^{-11}$ & 0.16 \\
\hline
\hline
\l{drake}
\end{tabular}
\caption{Spontaneous mutation rates for various organisms taken
from \cite{Drake98}. The first two organisms have RNA as genetic material 
and the rest are DNA based.}
\end{table}

In the next subsection, we will discuss models in which \emph{mutations} occur 
either as copying errors in the genetic material during cell division or 
induced by some external influences. In Table \ref{drake}, the spontaneous 
mutation rates for some organisms are shown. They differ by orders of 
magnitude between RNA-based viruses whose mutation rate
per genome exceeds unity, and DNA-based organisms, which can afford 
the complex replication machinery needed to reduce the mutation rate to a 
much lower level. 
It has been suggested that the high mutation rate of RNA viruses, rather
than being due to the lack of correction mechanisms, may constitute
an adaptation to the rapidly fluctuating environments that these
organisms encounter (see the chapter by E. L\'azaro).   
Within the DNA organisms, the mutation rate per base is seen to 
decrease with increasing sequence length, and the 
mutation rate per genome is roughly constant for similar organisms. 
However, mutation rates per genome in higher eukaryotes become comparable to 
those of DNA-based microbes if referred to 
the \textit{effective} genome size, which excludes non-coding regions 
\cite{Drake98}.  

\begin{figure}
\begin{center}
\mbox{
\includegraphics[width=5.2cm,angle=0]{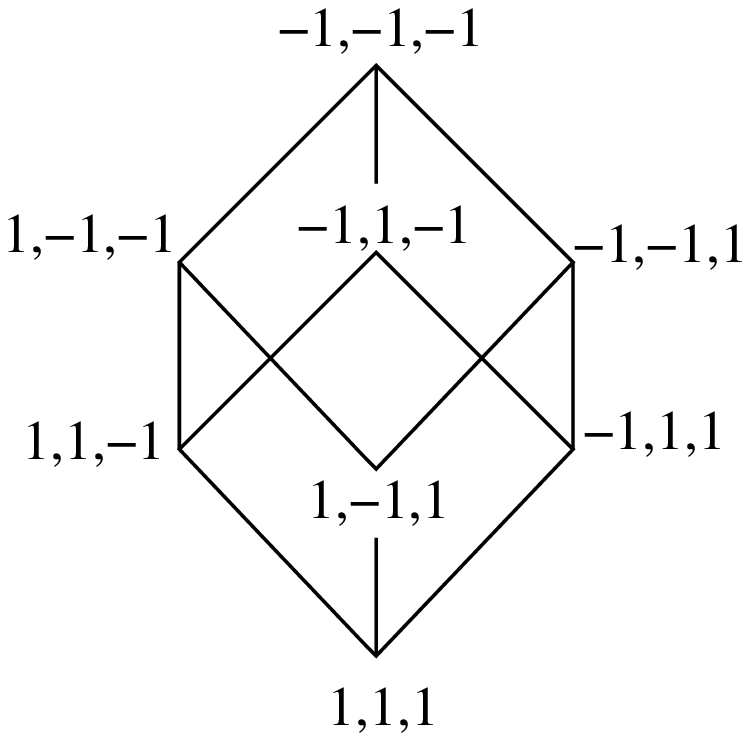}
\includegraphics[width=6cm,angle=0]{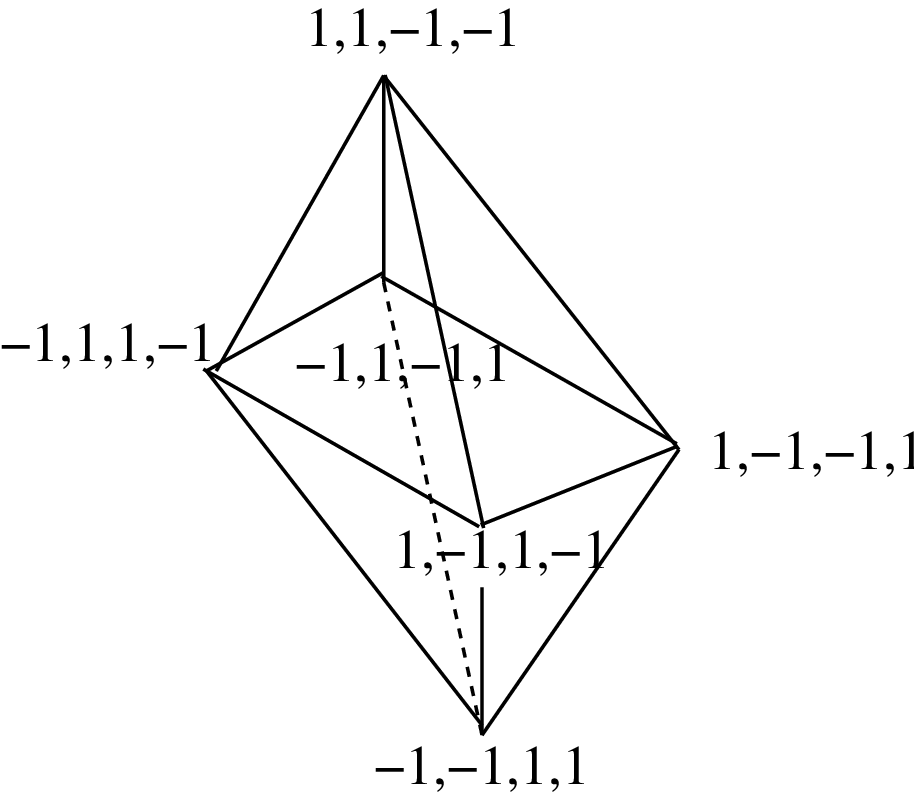}}
\caption{Examples of sequence spaces. Left panel:
Hamming space of binary sequences 
of length $N=3$. Right panel: 
Graph bipartitioning problem space for $N=4$. In both cases 
$\sigma_i = \pm 1$ and nearest neighbors are connected by lines.}
\label{space}
\end{center}
\end{figure}

Before we describe the mutation-selection models, we need to specify the 
space of sequences on which the evolutionary dynamics operate. 
The structure and geometry of the sequence
space depends on the nature of the allowed moves that change one 
sequence into another. 
In the simplest case of a genome of fixed length $N$ subject only to 
point mutations (which we will restrict ourselves to throughout this chapter), 
the natural choice for the sequence space is the Hamming space with 
$\ell^{N}$ points. Two sequences $\sigma$ and $\sigma'$ are separated 
by the Hamming distance $d(\sigma,\sigma')$ which is given by
\be
d(\sigma,\sigma^{\prime})= \sum_{i=1}^{N} 
(1- \delta_{\sigma_{i},\sigma_{i}'})  .
\ee
The Hamming distance simply counts the number of letters in which the two 
sequences differ, that is, the number of point mutations needed to mutate 
$\sigma$ into $\sigma'$ (and viceversa). The Hamming space for $N = 3$ 
and $\ell=2$ is shown in Fig.~\ref{space} (left). 
The sequences are located at
the corners of a cube, which for general $N$ becomes the $N$-dimensional
\textit{hypercube}. 

To give an example of a sequence space with a somewhat different geometry, 
we consider the Graph Bipartitioning Problem (GBP) \cite{Fu86}
(see also Sect.~\ref{correlated}).  
In the GBP, as the name suggests, the problem 
is to partition a graph with given connections into two sets $A$ and $B$
with equal number of vertices,  
such that the number of connections between $A$ and $B$ is minimised. 
A bipartitioning configuration is mapped onto a binary sequence by setting
$\sigma_{i}=1$ if the vertex $i$ belongs to set $A$, and  $\sigma_i = -1$ 
else. 
Thus the sequence space consists of those  
$N \choose N/2$ configurations $\sigma$ for which 
$\sum_{i} \sigma_{i}=0$, a subset of the Hamming space.
An elementary move exchanges a pair of vertices between the sets 
$A$ and $B$.  
Two configurations are said to be at a distance 
$d_{\mathrm{GBP}}(\sigma, \sigma') = d$ if they can be related by $d$ 
exchange moves, so 
that $d_{\mathrm{GBP}}$ is half of the Hamming distance defined 
above. The GBP sequence space for $N=4$ is 
shown in Fig.~\ref{space} (right).   

The Hamming space as well as the GBP sequence space are symmetric and regular 
graphs, in the
sense that each vertex has the same number of neighbors and all vertices are 
equivalent.
This is no longer true if mutations that change the sequence length through
deletions, insertions or gene duplications are taken into account.
Genetic recombination, which is of crucial importance for sexual reproduction,
leads to additional complications, because it introduces moves which involve
pairs of sequences \cite{Stadler02}.

We return to the case of point mutations acting on sequences of fixed 
length 
$N$, and proceed to derive an expression for the mutation probabilities taking
one sequence to another. 
If the mutations change 
a letter $\sigma_{i}$ to any one of the other $\ell -1$ values with a 
probability 
$\mu$, independent of the identity of the letter and the
other letters in the sequence, then the \textit{probability}
to mutate a sequence $\sigma'$ to $\sigma$ can be written as
\be
p(\sigma' \rightarrow \sigma)= 
\left( \frac{\mu}{(\ell-1)(1-\mu)}\right)^{d(\sigma,\sigma')} (1-\mu)^{N}. 
\l{mutprob1}
\ee
Obviously, this probability is the same for all $\alpha_{d}$ 
sequences which are at a constant Hamming distance $d$ from sequence 
$\sigma$, where $\alpha_d$ is given by  
\be
\alpha_{d}={N \choose d}  (\ell-1)^{d}  \l{alphak}.
\ee
This can be seen by noting that there are $N \choose d$ ways of choosing 
$d$ letters at which a sequence differs from $\sigma$ and each of 
these $d$ letters can take $\ell-1$ values. For large $N$, 
most of the sequences are located in a belt of width $\sim \sqrt{N}$ 
around the 
distance $d_{\mathrm{max}}=N (\ell -1)/\ell$ away from $\sigma$. Using 
(\ref{alphak}), 
it is easily checked that $\sum_{\sigma} p(\sigma' \rightarrow \sigma)=1$.

Similar to the transition from Wrightian to Malthusian fitness, 
in the continuous time limit 
the mutation probability (\ref{mutprob1}) has to be 
replaced by the \textit{mutation rate} $\gamma(\sigma' \to \sigma)$, such that
for generation time $\Delta t \to 0$
\be
\l{Contmut}
p(\sigma' \to \sigma) \approx \delta_{\sigma', \sigma} + \Delta t \; 
\gamma(\sigma' \to \sigma).
\ee
Denoting the mutation rate 
per letter by $\tilde \mu$ and setting $\mu = \tilde \mu \, \Delta t$ 
in (\ref{mutprob1}) yields 
\be
\gamma(\sigma' \rightarrow \sigma) = \cases 
{0 & {$: \;\;d(\sigma',\sigma) > 1 $} \cr
\tilde \mu/(\ell -1) & {$: \;\;d(\sigma',\sigma)=1 $} \cr
-N \tilde \mu & {$:\;\;d(\sigma',\sigma)=0 $}}.  \l{mutprob2}
\ee
The normalization condition for mutation rates reads
$\sum_{\sigma} \gamma(\sigma' \rightarrow \sigma) = 0$. 


\subsection{Mutation-selection models}
\l{muse}

We now discuss models of adaptation that incorporate the two competing 
processes discussed above, namely, mutation and selection. While mutation 
increases genetic 
diversity, selection tends to contain the population at fit sequences. In 
case selection wins out, 
one obtains a population in which individuals are genetically closely related 
else a heterogeneous 
population distributed over the entire sequence space results. 
In this article, we will mainly discuss the so-called \textit{coupled models} 
in which 
the mutations occur only during replication. In the \textit{paramuse models}, 
on the 
other hand, mutation and selection occurs in parallel, and they will be 
discussed 
here briefly. We refer the reader for more details to the reviews 
\cite{Baake00,Baake01} and references therein. 
While one may expect both types of mutation mechanisms to be relevant in 
describing evolution, the jury is still out on their relative importance. 
For this reason, both classes of models have been analysed in detail 
and the relationship between them has been explored, with regard
to both static \cite{Wiehe95} and dynamic \cite{Saakian04} properties.  

The models discussed below work under the following two assumptions:

\begin{itemize}

\item[(i)] \textbf{Infinite population}, i.e., the total 
population size $M \gg \ell^{N}$, the total number of genotypes available. 
Under this
assumption a 
deterministic description suffices and we can write down the time evolution 
equation 
for the average population fraction $X(\sigma,t)$ of sequence $\sigma$ at 
time $t$.  
Although this is often unrealistic,  
the analysis is simpler in this limit which in many cases can be adapted 
to the finite population case to provide quantitative agreement with  
experiments \cite{Domingo78,Wahl00,Rouzine03}. The infinite
population limit can be justified if the population is known to be localized
in a small region of sequence space
around a fitness peak, if one is interested in a short piece of the genome 
such as a single regulatory binding
site \cite{Gerland02} (see also Sect.~\ref{Modified}) 
or if one works in the population genetics setting, where
the letters in the sequence are alleles of a gene, rather than single 
nucleotides.   

\item[(ii)] \textbf{Asexual reproduction} which dominates in the lower forms 
of life 
such as virus and bacteria, and digital organisms. We will mainly consider 
haploid 
organisms but diploids are briefly discussed in Section \ref{extensions}. 
However, we do not 
consider the case of sexual reproduction; a comparison between 
sexual and asexual reproduction modes in the context of sequence space
models can be found in \cite{Higgs94}.

\end{itemize}


\subsubsection{Paramuse models}

In the paramuse models, introduced by Crow and Kimura \cite{Crow70}, one  
assumes error-free replication and 
mutations are induced by the environment through radiation, thermal 
fluctuations etc. \cite{Baake00}. The equation for the rate of change 
$\dot{X}(\sigma,t) = \partial X(\sigma,t)/\partial t$ of the 
fraction $X(\sigma,t)$ of the population with sequence $\sigma$ is given by 
\be
\dot{X}(\sigma,t)=
[w(\sigma)-\sum_{\sigma'} w(\sigma') X(\sigma',t)] X(\sigma,t)+
\sum_{\sigma'} \gamma(\sigma' \rightarrow \sigma) X(\sigma',t). 
\l{ckxc}
\ee
The first term is the selection term while the contribution from the mutations 
is contained in the last term. 
The evolution equation (\ref{ckxc}) is nonlinear in $X(\sigma,t)$ due to 
the second term on the right hand side, which is required to ensure
the normalisation $\sum_{\sigma} X(\sigma,t) = 1$. 
This nonlinearity can be eliminated by passing
to unnormalised population variables $Z(\sigma,t)$ defined by 
\be
Z(\sigma,t)= X(\sigma,t) \;\mbox{exp} \left[ \sum_{\sigma{'} }  w(\sigma{'}) 
\int_{0}^{t} d \tau X(\sigma{'},\tau) \right] \l{normzxc}
\ee
which satisfy the linear equation \cite{Baake97b}
\be
\frac{\partial Z(\sigma,t)}{\partial t}=w(\sigma) Z(\sigma,t)+
\sum_{\sigma'} \gamma(\sigma' \rightarrow \sigma) Z(\sigma',t). 
\l{ckzc}
\ee
Equation (\ref{ckxc}) follows from (\ref{ckzc}) using the relation 
\be
X(\sigma,t)=\frac{Z(\sigma,t)}{\sum_{\sigma'} Z(\sigma',t)}. \l{normxz}
\ee 
Inserting the explicit form (\ref{mutprob2}) for the 
mutation rates, it can be shown that 
the vector ${\bf Z} (t)=(Z(\sigma^{1},t),...,Z(\sigma^{S},t))$, where
the index labels the $S = \ell^N$ points in sequence space, obeys 
a Schr\"odinger equation in imaginary time 
\be
\frac{\partial {\bf Z}(t)}{\partial t}=H {\bf Z}(t)
\l{spinchain}
\ee
with quantum spin Hamiltonian $H$ in one dimension. 
Specifically, for $\ell=2$, one obtains the Hamiltonian of 
an Ising chain in the presence of a 
transverse magnetic field (mutations) with general interactions 
(specified by the fitness landscape) \cite{Baake97b}; for an explicit
example see Eq.~(\ref{Qspin}).
This model has been solved exactly for a variety of fitness landscapes
using methods of quantum  
statistical physics \cite{Saakian04,Baake97b,Wagner98}. A similar analysis 
has also 
been carried out for the 
biologically relevant case of $\ell=4$ \cite{Hermisson01}.  

\subsubsection{Coupled (quasispecies) dynamics}

In the quasispecies model introduced by Eigen in the context of 
prebiotic evolution
\cite{Eigen71,Eigen77,Eigen88}, the mutations are copying errors that 
occur during the reproduction process. This implies that the population 
fraction $X(\sigma,t)$ evolves according to 
\be
\dot{X}(\sigma,t)=\sum_{\sigma{'}} p(\sigma{'} \rightarrow \sigma) 
W(\sigma{'}) 
X(\sigma{'},t)-\left(\sum_{\sigma{'}} W(\sigma{'}) X(\sigma{'},t) \right) 
X(\sigma,t) \l{exc}
\ee
which can be linearised by a transformation analogous to (\ref{normzxc}) to 
yield the linear equation 
\be
\dot{Z}(\sigma,t)=
\sum_{\sigma{'}} p(\sigma{'} \rightarrow \sigma) W(\sigma{'}) Z(\sigma{'},t). 
\ee
In discrete time this model takes the form 
\be
X(\sigma,t+1)=\frac{\sum_{\sigma{'}} W(\sigma{'}) 
p(\sigma{'} \rightarrow \sigma) X(\sigma{'},t)}
{\sum_{\sigma{'}} W(\sigma{'}) X(\sigma{'},t) } \l{exdis}
\ee
where the denominator arises due to the normalisation. The discrete time 
analog of the transformation (\ref{normzxc}) is given by
\be
\l{Eigendisc}
Z(\sigma,t)=X(\sigma,t) \; \prod_{\tau=0}^{t-1} \sum_{\sigma{'}} W(\sigma{'}) 
X(\sigma{'},\tau)
\ee
As before, the unnormalised variables obey a linear equation given by 
\be
Z(\sigma,t+1)=\sum_{\sigma{'}} p(\sigma{'} 
\rightarrow \sigma) W(\sigma{'}) Z(\sigma{'},t). \l{ezdis}
\ee

The use of the Wrightian (discrete time) fitness $W(\sigma)$ in the continuous
 time equation 
(\ref{exc}) requires some explanation. First, it ensures that the stationary 
solutions of 
(\ref{exc}) and (\ref{exdis}) are identical. Second, it reflects the fact that 
(\ref{exc}) is invariant (up to a rescaling of time) under multiplication of 
the fitnesses
by a constant factor,
$W(\sigma) \to C W(\sigma)$, which is an exact symmetry of the discrete time 
equation 
(\ref{exdis}), whereas the continuous time paramuse dynamics (\ref{ckxc}) is 
invariant
under additive shifts $w(\sigma) \to w(\sigma) + C$ \cite{Baake00,Wiehe95}.  
In fact, (\ref{exc}) is \textit{not}
the continuous time limit of (\ref{exdis}). Instead, inserting 
(\ref{Malthus}) and (\ref{Contmut}) 
in (\ref{exdis}) and taking $\Delta t \to 0$, one
obtains the paramuse dynamics (\ref{ckxc}).  
In this sense (\ref{exc}) is somewhat intermediate between the discrete
time model (\ref{exdis}) and the continuous time dynamics (\ref{ckxc}).

For the discrete time model (\ref{exdis}) one can represent the 
evolutionary histories as configurations on a  
two-dimensional lattice with the two axes directed along the sequence and 
along 
time, with a spin variable $\sigma_{i}(t)$ at each site. 
Writing the evolution equation (\ref{ezdis}) 
for the vector ${\bf Z}(t)$ in the form 
\be
{\bf Z}(t+1)=T_{t+1,t} \;{\bf Z}(t)
\ee
then suggests to interpret $T_{t+1,t}$ as the transfer matrix 
of a two-dimensional classical spin model which relates the 
probability of a configuration in one row of the lattice to the next one 
\cite{Kogut79}. 
For $\ell=2$, this $2^{N} \times 2^{N}$ matrix can be written (up to 
a multiplicative constant) as
\be
T_{t+1,t}[\{\sigma_{i}(t+1)\}, \{\sigma_{i}(t)\}] =
\mbox{exp}[\ln W(\{\sigma_{i}(t)\}) + J 
\sum_{i=1}^{N} \sigma_{i}(t+1) \sigma_{i}(t)] \l{transfer}
\ee
where 
\be
J = \frac{1}{2} \ln( \mu^{-1} - 1).
\l{beta}
\ee 
Thus $T_{t+1,t} = \exp[- \tilde H]$, where $\tilde H$ is the 
Hamiltonian of a two-dimensional 
Ising model\footnote{The Ising Hamiltonian $\tilde H$ should not be confused
with the Hamiltonian $H$ of the quantum spin chain in (\ref{spinchain}).} 
with nearest neighbor interactions of strength $J$ 
along the time direction and general interactions [determined by the 
fitness landscape $W(\sigma)$] along the sequence direction
\cite{Leuthausser86}. 
The expression (\ref{beta}) shows that the interactions along the time 
direction are ferromagnetic (antiferromagnetic) whenever 
$\mu < 1/2$ ($\mu > 1/2$), while 
for $\mu = 1/2$ the 
sequence is completely randomized in each time step and the interaction
vanishes. 

Clearly, to obtain the distribution of sequences at time slice $t$, 
one needs to solve iteratively for all the $t-1$ preceding layers. In the 
steady state for which  $t \ra \infty$ one requires the properties of 
the last ``surface'' layer  coupled to a semi-infinite ``bulk''.
Since the transfer matrix (\ref{transfer}) does not contain any couplings
along the sequence direction in the last layer $t+1$, the boundary condition 
for this semi-infinite spin model corresponds to a free surface 
\cite{Tarazona92}.   

\section{Simple fitness landscapes}
\label{Simple}

So far we have discussed the general equations governing the evolution of a 
population 
with mutations, but the fitness landscape was not 
specified.
We do so now and begin with landscapes that are ``simple" in that the fitness 
depends only on the distance from a given (\textit{master}) 
sequence, which is usually 
the genotype of highest fitness\footnote{In the context of population genetics,
the master genotype is often referred to as the 
\textit{wild type}.}. 
Such landscapes are called \textit{permutation invariant}, because
the fitness depends only on the number of mismatches relative to the master
sequence, but not on their position.
Using this symmetry, the $\ell^{N}$ population variables can be grouped
into $N+1$ \textit{error classes}, which greatly facilitates both numerical
and analytic work \cite{Nowak89}.

\subsection{The error threshold: Preliminary considerations}

Much of 
this section is devoted to a discussion of the error threshold phenomenon, 
which 
refers to the loss of genetic integrity when mutations are increased beyond a 
certain threshold. We consider only the stationary population distribution
which is established after a long time. The linearity of both the continuous
and discrete time evolution equations (\ref{ckzc}, \ref{exdis}, \ref{ezdis}) 
implies that the stationary distribution
is identical to the \textit{principal eigenvector} of the matrix multiplying
the population vector on the right hand side, i.e., the eigenvector with
the largest eigenvalue. The principal eigenvalue is related to the mean 
population fitness in the stationary state. In this sense, 
the analysis of different fitness landscapes and mutation schemes is reduced 
to the 
investigation of the spectral properties of the corresponding evolution 
matrices \cite{Rumschitzki87}.  

The error threshold separates two regimes of mutation-selection balance  
characterized by a qualitatively different structure of the principal
eigenvector. For small mutation rates the eigenvector is localized around
the master sequence, i.e. only the entries corresponding to the 
dominant genotype and a few of its nearby mutants carry appreciable weight.
Following Eigen and Schuster \cite{Eigen77}, such a localized population
distribution is referred to as a \textit{quasispecies}. When the mutation
rate is increased beyond the error threshold, the principal eigenvector
becomes delocalized and the population spreads uniformly throughout the
sequence space. In this regime finite population effects, which are neglected
in the models considered here, become extremely important: Rather than 
covering the entire sequence space, which is impossible given the vast
number of sequences, a finite population forms a localized cloud which
wanders about randomly \cite{Derrida91}.

Since the eigenvectors and eigenvalues of any finite matrix depend smoothly
on its entries, the error threshold can become \textit{sharp}, in the 
sense of being associated with some non-analytic behavior of the population
distribution or the mean population fitness, only in the limit $N \to \infty$.
We shall see below that in order to maintain the localized quasispecies in this
limit, it is usually necessary to either reduce the single site mutation 
probability $\mu$, 
such that the mutation probability per genome 
$\mu N$ remains constant, or to increase the
selective advantage of the master sequence with increasing $N$.

\subsection{Error threshold in the sharp peak landscape}
\l{Sharppeaksection}

We demonstrate the error threshold in the case of a single sharp peak 
landscape which is defined as 
\be
W(\sigma)=W_{0} \delta_{\sigma,\sigma_{0}}+ (1-\delta_{\sigma,\sigma_{0}}) 
, \;\;\;\; W_{0} > 1.  \l{spl}
\ee
Here $\sigma_{0}$ denotes the master sequence, and $W_0$ is the selective 
advantage of the master sequence
relative to the other sequences, whose Wrightian fitness has been normalized 
to unity.  
We anticipate the error threshold to occur
for $\mu \rightarrow 0$, $N \rightarrow \infty$, keeping the mutation
rate per genome $\mu N$ finite. 
Let us consider the coupled model in discrete time\footnote{Recall that in 
the steady state, both versions of the coupled model are identical.} 
defined by 
(\ref{exdis}) with the choice (\ref{spl}). 
In the limit $\mu \to 0$ the mutations taking the mutants back into the
master sequence can be neglected\footnote{Neglecting back mutations
towards the master sequence is common in population genetics,
where it is referred to as a \textit{unidirectional} mutation scheme 
\cite{Baake01}.
It simplifies the analytic treatment \cite{Higgs94,Wiehe97}, and will be used
repeatedly in this article as an approximation which is expected to become 
exact for $\mu \to 0$, $N \to \infty$.}, 
and the only nonzero contribution to $X(\sigma_0)$  on the right hand
side of (\ref{exdis}) is that for $\sigma{'}=\sigma$. This yields
\be
\l{x0eigen}
X(\sigma_{0})= \frac{W_{0} e^{-\mu N}-1}{W_{0}-1}
\ee
which is an acceptable solution provided $\mu N \leq \ln W_{0}$. Thus, a 
phase transition occurs at the critical mutation probability 
\be
\mu_{c}=  \frac{\ln W_{0}}{N}  \l{etsplc}
\ee
beyond which the population cannot be maintained at
the peak of the landscape. Close to $\mu_{c}$, the fraction of population at 
the master sequence behaves as 
\be
X(\sigma_{0}) \approx \frac{N}{W_{0}-1} (\mu_{c}-\mu) \l{xspl}
\ee
thus approaching zero continuously at $\mu_{c}$. 
The above results are also confirmed by a detailed numerical analysis for 
finite $\mu$ and $N$, 
in which the population was grouped into error classes at constant Hamming 
distance from the master sequence
and the population in the error classes as well as the eigenvalues of the 
evolution matrix
were followed as a function of $\mu$ \cite{Nowak89}. 

\begin{figure}
\begin{center}
\includegraphics[width=8cm,angle=0]{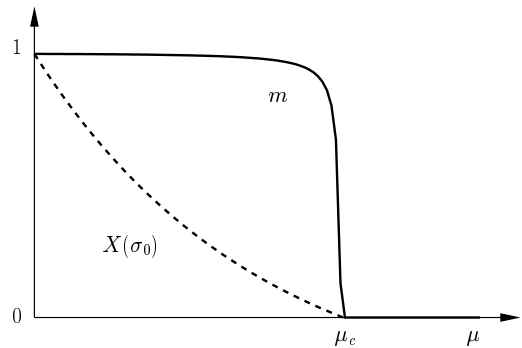}
\caption{Figure to show the continuous transition in the 
fraction $X(\sigma_{0})$ of the master 
sequence and the (almost) 
discontinuous one in the overlap $m$ as a function of mutation 
rate $\mu$, for $N=1000$ and $W_{0}=4$.}
\label{splfig}
\end{center}
\end{figure}

The way in which the error threshold condition (\ref{etsplc}) combines
mutation rate, sequence length and selective advantage is the
central result of quasispecies theory. In particular, it shows that in order
to maintain a localized quasispecies at finite single site mutation rate in 
the limit $N \to \infty$, the selective advantage has to increase 
\textit{exponentially} with $N$ \cite{Franz97}. Under the assumption that
typical selective advantages do not depend strongly on sequence length,
Eq.~(\ref{etsplc}) also provides some rationalization for the observation
that the product $\mu N$ is roughly constant within classes of similar
organisms (see Sect.~\ref{basic}). 
On the other hand, at given achievable values of the 
replication accuracy and the selective advantage, the condition
$\mu < \mu_c$ place an upper bound $N_\mathrm{max}$ on the sequence
length, beyond which genetic integrity is lost. Elsewhere in this
book Ester L\'azaro presents substantial
evidence that RNA viruses have evolved to reside close to this
threshold, possibly because this allows them to maintain a maximal
genetic variability which is needed to rapidly adapt to 
changing environments (see also Sect.~\ref{RNAviruses}). 

Neglecting back mutations to the master sequence 
allows to derive an expression for the
mean Hamming distance
to the master sequence, which reads \cite{Wiehe97}
\be
\l{dmean}
\langle d(\sigma,\sigma_0) \rangle = \frac{W_0 N \mu}{W_0 e^{-N \mu} - 1}.
\ee
The mean Hamming distance
is finite for $\mu < \mu_c$ and diverges as $(\mu_c - \mu)^{-1}$
as the error threshold is approached.  
This provides an alternative characterisation of the threshold.
A related quantity, which has been proposed as an order parameter
for the transition, is the \textit{mean overlap} 
\be
\l{overlap}
m = 1 - \frac{2 \langle d(\sigma,\sigma_0) \rangle}{N}
\ee 
between
the master sequence and a randomly chosen sequence \cite{Franz97}.
Since $\langle d (\sigma,\sigma_0) \rangle$ remains finite for 
$N \to \infty$ in the localised phase, the overlap is $m = 1$
in this limit and jumps discontinuously to $m = 0$ at the threshold.
Figure~\ref{splfig} displaying the two order parameters considered in the 
above 
discussion illustrates that the nature of the transition --  
continuous or discontinuous -- depends to some extent on the quantity
under consideration\footnote{In contradiction to the discussion 
above, a numerical 
study based on the mapping to a 
two-dimensional Ising model described in Sect.~\ref{muse} deduced that both 
$m$ and $X(\sigma_{0})$ change smoothly at the transition 
\cite{Tarazona92}. However in this study, a scaling analysis with 
genome length (akin 
to finite size scaling analysis in statistical mechanics) was not carried 
out to obtain the behavior in the limit $N \to \infty$.}.

Yet another characterization of the error threshold relies on the notion of the
\textit{consensus sequence} $\sigma^c$, which carries at each site $i$
that letter $\sigma_i^c$ which is most frequently represented in the population.
It is easy to see that, for symmetry reasons, the consensus sequence in
the sharp peak landscape (\ref{spl}) coincides with the master sequence, $\sigma^c = \sigma_0$, 
throughout the localized phase; this is true for general permutation-invariant
single peak landscapes. In the delocalized phase, where the population is uniformly spread throughout
sequence space for $N \to \infty$, all letters appear with equal probability and
the consensus sequence cannot be defined. 
This is an artifact of the assumption of infinite population
size: a finite population retains some genetic structure even in 
a flat fitness landscape and diffuses through sequence space as a cloud
centered around a moving consensus sequence $\sigma^c(t)$ \cite{Derrida91}. 
Thus at the error threshold the consensus sequence ceases to be pinned
to the master sequence and becomes time-dependent. This
criterion to locate the transition is particularly useful in complex fitness landscapes,   
where the most-fit master sequence is not known \cite{Bonhoeffer93} (see Sect.\ref{Complex}).
Similarly, in experimental studies of microbial populations such as RNA viruses, the consensus
sequence is taken to represent the (unknown) wildtype genome, and the genetic spread of the population around
$\sigma^c$ is interpreted as a measure of the balance between mutational and
selective forces (see the chapter by E. L\'azaro).  
  
\subsection{Exact solution of a sharp peak model}

A variant of Eigen's model was solved exactly for any $N$ 
in \cite{Galluccio97}. The model 
is defined in discrete time but the mutations are restricted
to mutants within Hamming distance equal to one, as for the continuous
time mutation rates (\ref{mutprob2}). In addition, mutations 
are assumed to occur in the whole 
population \emph{before} the reproduction process. With the fitness landscape
(\ref{spl}) this leads to the linear evolution equation 
\bea
Z(\sigma,t+1)=[1+(W_{0}-1) \delta_{\sigma,\sigma_{0}}] \times
\hspace*{5.cm}    \\
\times \left[ (1-N \mu) Z(\sigma,t)+ 
\mu \sum_{\sigma{'}} Z(\sigma{'},t) \; \delta_{d(\sigma',\sigma),1}\right]
\nonumber
\l{galmodel}
\eea
for the unnormalised population variables. Note that the model is well defined
only for $N \mu < 1$.

At large times, $Z(\sigma,t+1) \approx \Lambda Z(\sigma,t)$ where 
$\Lambda$ is the largest eigenvalue of the evolution matrix on the right hand
side of (\ref{galmodel}). 
In the delocalised phase, the population is spread over the entire sequence
space with mean fitness $W = 1$,  
so that $\Lambda=1$ 
whereas in the localised phase, a finite fraction has fitness $W_0 > 1$ 
and hence $\Lambda > 1$. For any 
$N$, the eigenvalue $\Lambda$ is determined by the exact equation
\be
\frac{W_{0}}{W_{0}-1}=\frac{1}{2^N} \sum_{k=0}^{N} {N \choose k} 
\frac{\Lambda}{\Lambda-1+2k \mu}. 
\l{eigen}
\ee
Due to the $k=0$ term on the RHS of the above equation, it is evident 
that $\Lambda$ can take a value equal to 
$1$ only in the $N \to \infty$ limit. Thus, there is no phase transition for 
any finite $N$. 

In the limit $N \to \infty$, $\mu \to 0$ with $N \mu < 1$ fixed the 
eigenvalue is given by the expression
\be
\l{eigenvalue}
\Lambda = \max \{ 1, W_0 (1 - N \mu) \},
\ee
which sticks to unity beyond the critical mutation strength
\be
\l{Galmuc}
\mu_c = \frac{W_0-1}{W_0 N}.
\ee
Incidentally, the above expression for $\mu_{c}$ can be obtained 
using (\ref{x0eigen}) by expanding the exponential to first order in 
$\mu N$. This is required to ensure that $\mu N < 1$ is satisfied for 
any $W_{0} > 1$. In both cases
the selective advantage needed to localize the quasispecies is the inverse
of the \textit{copying fidelity}, i.e. the probability of creating an 
error-free offspring. 

The behavior of other quantities at the threshold follow from that of 
$\Lambda$. For example, the fraction of the population residing at the
master sequence is given by 
\be
\l{galx0}
X(\sigma_0) = \frac{W_0 (\Lambda - 1)}{(W_0 - 1) \Lambda}
\ee
which vanishes linearly in $\mu_c - \mu$ at the threshold, 
and the mean Hamming distance from the master sequence is 
\be
\l{gald}
\langle d(\sigma,\sigma_0) \rangle = \frac{N \mu}{\Lambda - 1}
\ee
which diverges as $(\mu_c - \mu)^{-1}$. The expressions 
(\ref{Galmuc} - \ref{gald})
are valid in the asymptotic limit $N \to \infty$, but systematic expansions 
of these quantities in powers of $1/N$ are also available \cite{Galluccio97}.

Comparing the expressions (\ref{xspl}) and (\ref{dmean}) to 
(\ref{galx0}) and (\ref{gald}) respectively, we see that 
$X(\sigma_0)$ and $\langle d(\sigma,\sigma_0)
\rangle$ behave qualitatively similar in the two models as the error threshold
is approached. This is a simple example of the 
principle of \textit{universality} commonly encountered at physical phase
transitions, which states that the way in which singular quantities vanish
or diverge at the transition is independent of detailed properties of the 
model.
\subsection{Modifying the shape of the fitness peak}
\l{Modified}

Since the sharp peak landscape (\ref{spl}) was chosen for its
simplicity, and not because it is expected to be biologically realistic,
it is important to investigate how the error threshold 
phenomenology depends on the shape of the fitness peak. In this section
we discuss some illustrative examples. A method for solving the 
stationary quasispecies equation for general peak shapes has been developed
by Peliti \cite{Peliti02}. It employs a strong selection limit, in which
the fitness is written as $W(\sigma) = \exp[N \Phi(\sigma)]$ and 
the limit $N \to \infty$ is carried out at fixed mutation probability $\mu$.

\subsubsection{Peak height versus peak width}

We first consider a landscape 
with one sharp global maximum and a broad peak of lower fitness 
separated by a flat landscape. This is defined as
\be
W(\sigma)=W_{0} \delta_{\sigma,\sigma_{0}}+ W_{N} \delta_{\sigma,\sigma^{N}}+
W_{N-1} \delta_{d(\sigma,\sigma^{N}),1}+\sum_{j \neq 0,N-1,N} 
\delta_{d(\sigma,\sigma_0),j}
\l{twopeaks}
\ee
where $\sigma^{N}$ is the sequence at maximal 
Hamming distance $N$ from $\sigma_{0}$ and $W_{0} > W_{N} > W_{N-1} > 1$.
By placing the two fitness peaks at the two poles $\sigma_0$ and 
$\sigma^{N}$ of the sequence space, the permutation symmetry of the landscape
is preserved and the population can be subdivided into error classes.    
The coupled model with the landscape (\ref{twopeaks}) has been studied in both 
continuous \cite{Schuster88} and 
discrete time \cite{Tarazona92}. Interestingly, with increasing mutation rate, 
the quasispecies shifts abruptly from the sequence $\sigma_{0}$ to the 
broader peak around $\sigma^{N}$ finally delocalising over the whole 
sequence space. For large mutation rates, the quasispecies
is more comfortable at the lower peak surrounded by an extended 
region of elevated fitness than at the (globally optimal) isolated
master sequence.  


\subsubsection{Mesa landscapes}

Broad fitness peaks arise naturally in the evolution of regulatory binding 
sites 
\cite{Gerland02,Peng03,Berg04}. In this context the fitness of a given 
regulatory sequence
can be plausibly related to the binding probability of the corresponding 
transcription factor. Simple thermodynamic models predict that the binding
probability depends on the number of mismatches $d(\sigma,\sigma_0)$ with 
respect to the regulatory master sequence $\sigma_0$ through a Fermi function, 
\be
\l{binding}
p_b(d) = \frac{1}{1 + \exp[\epsilon(d - d_0)/k_{\mathrm B} T]},
\ee
where $\epsilon$ is the binding energy per mismatch, $\epsilon d_0$ is the 
chemical
potential corresponding to the concentration of the transcription factor, and 
$k_{\mathrm B} T$ is the thermal energy at temperature $T$. For 
$\epsilon/ k_{\mathrm B} T
\gg 1$ the binding probability drops abruptly from $p_b = 1$ to 
$p_b = 0$ when $d$ exceeds the number $d_0$ of tolerable mismatches;  
a typical value of this ratio is $\epsilon/ k_{\mathrm B} T \approx 2$.

In the simplest scenario, the selective advantage of a regulatory sequence is 
assumed
to be proportional to the binding probability. This leads to a 
\textit{mesa-shaped} fitness landscape,
with a plateau of constant fitness and radius $d_0$ around the master 
sequence. 
In \cite{Gerland02} a detailed study of the error threshold in this landscape
was presented for continuous time paramuse dynamics with fitness landscape
$w(d) = w_0 p_b(d)$. An exact solution is possible
in a limit where $d$ becomes a continuous variable and the Fermi function 
(\ref{binding})
is replaced by a step function. Provided $d_0 \ll N$, 
the error threshold is found to take place at 
a critical mutation strength $\mu_c$ given by 
\be
\l{mesaet}
\mu_c = \frac{2 w_0}{N (1 + \eta^2/d_0^2)},
\ee
where $\eta$ is a constant of order unity. The critical mutation strength
is seen to increase with increasing $d_0$, illustrating the enhanced stability
of the quasispecies with increasing width of the fitness peak.
In the localized phase, the majority of the population is located near
the mesa edge at $d = d_0$, reflecting the exponential increase of the 
number (\ref{alphak}) of available genotypes with distance $d$. This 
is a purely entropic effect, which leads to a maximal \textit{fuzziness}
of regulatory motifs. 

Somewhat more realistically, one expects that the fitness depends not only
on the ability of the sequence to bind the transcription factor in a certain
cellular state, but also on its ability to \textit{avoid} binding in 
other states. This can be modeled by a fitness function which is proportional
to the difference between two Fermi functions (\ref{binding}) with 
different values of $d_0$, leading to a \textit{crater} landscape with a 
rim of high fitness around a fitness minimum at $d=0$ \cite{Berg04}. 
\subsubsection{Epistasis: Coupled dynamics}

Not all landscapes display the error threshold phenomenon.
We illustrate this point using 
the multiplicative (or Fujiyama) landscape as an example. In this case
\be
\l{Fuji}
W(\sigma)= \prod_{i=1}^{N} e^{\lambda \sigma_{i}} = 
\exp[\lambda(N - 2d(\sigma_0, \sigma))],
\ee
where 
for simplicity we choose $\ell=2$ and let $\sigma_{i}$ take values $\pm 1$.
For $\lambda > 0$ 
the master sequence is $\sigma_0 = (1,1,1...,1)$ and  
the Hamiltonian $\tilde H$ obtained from (\ref{transfer}) is 
\be
\tilde H= \sum_{i=1}^{N}[ -J \sigma_{i}(t+1) \sigma_{i}(t) - 
\lambda \sigma_{i}(t)].
\ee
Due to the absence of interactions along the sequence space direction, one 
obtains, for each position $i$, a one-dimensional Ising model 
in the presence of magnetic field $\lambda$. This  model 
is well known to lack a phase transition and due to the $\lambda$ term, the 
spins tend to align in the direction of the field. Correspondingly, 
a finite fraction of the population is maintained at the master
sequence for any value of the mutation rate. The full population
distribution has been worked out in \cite{Woodcock96}. 

In genetic terms, the multiplicative form (\ref{Fuji}) implies that the 
different gene loci contribute independently to the fitness, which is 
referred to as the absence of \textit{epistatic interactions}. In general,
one must distinguish between \textit{synergistic} or \textit{negative} 
epistasis, in which the (deleterious) effect of an additional
mutation \textit{increases}
with increasing distance from the wild type (master sequence), and 
\textit{diminishing returns} or \textit{positive} epistasis, when the
effects of mutations decreases with increasing distance\footnote{This 
nomenclature
is based on \cite{Baake01,Wiehe97}, but it 
does not appear to be unambiguous; in \cite{Kondrashov88} a definition of 
positive and negative epistasis is used which is opposite to the 
present one.}. 
The sharp peak landscape (\ref{spl}) is an extreme case of positive 
epistasis, because 
after the first mutation away from the master sequence, any additional
mutation does not affect the fitness at all. An extreme limit of negative 
epistasis is represented by the case of
truncation selection, where the Wrightian fitness vanishes beyond a 
critical Hamming distance $d_c$ \cite{Kondrashov88}. As we discuss 
below, whether or not an error threshold occurs depends on the 
behavior of the landscape at large Hamming distance from the master sequence.

\begin{figure}
\begin{center}
\includegraphics[width=11.cm,angle=0]{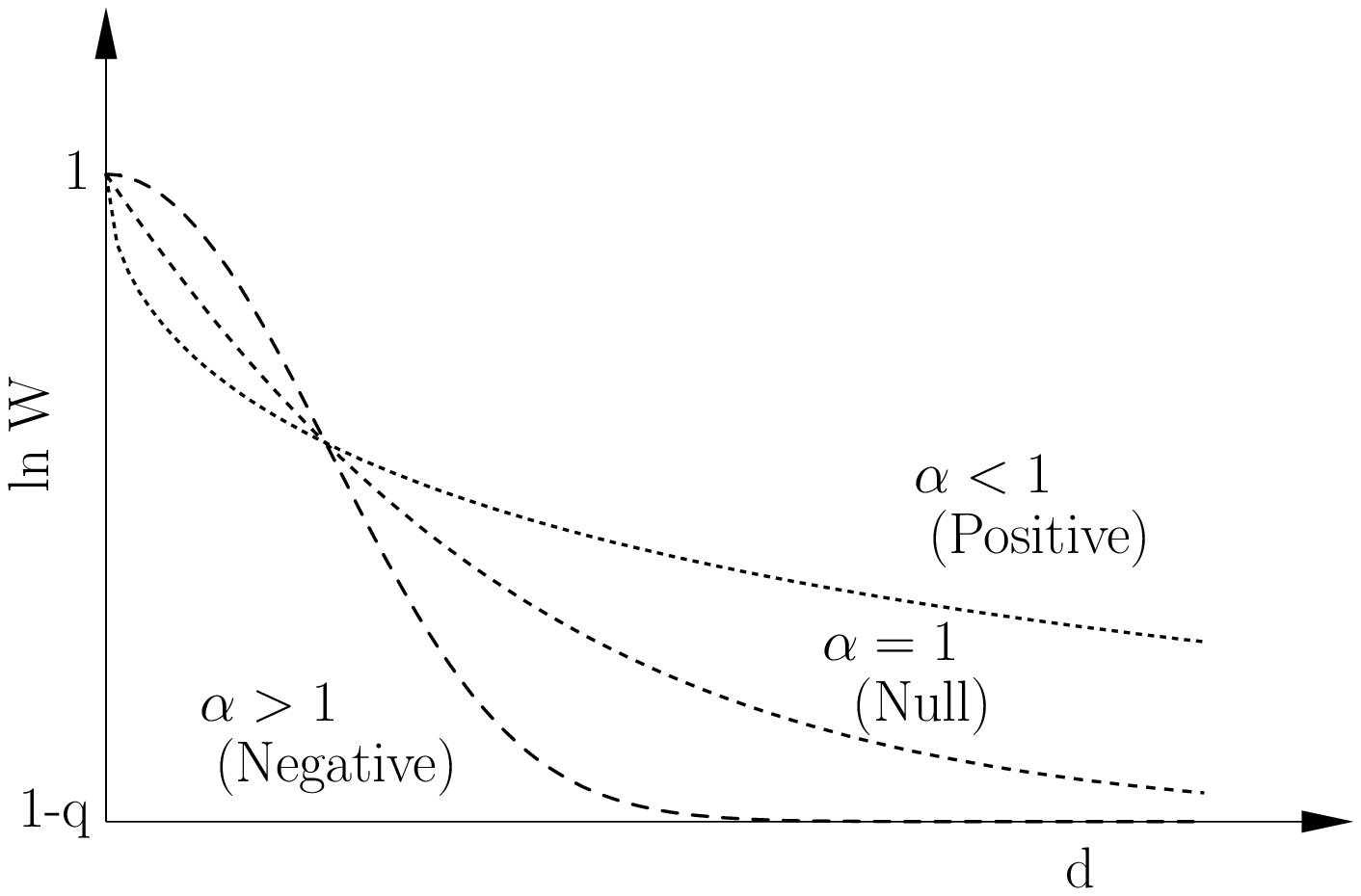}
\caption{Illustration of the fitness landscape (\ref{Wiehe1})
with $s = q = 0.5$ and three different values of $\alpha$.}
\label{epiland}
\end{center}
\end{figure}

Consider a general fitness landscape defined by \cite{Wiehe97}
\be
\l{Wiehe1}
W(\sigma) = q (1 - s)^{d(\sigma,\sigma_0)^{\alpha}} + 1 - q
\ee
where $0 \leq q, s \leq 1$ and $\alpha > 0$. Two cases need to be 
distinguished: for $q < 1$, the lower bound on the fitness is nonzero and 
when $s \to 1$ it becomes of sharp peak type (\ref{spl}) with the (relative)
selective advantage $1/(1 - q)$ for the master sequence, while 
for $q=1$, the multiplicative
form (\ref{Fuji}) with $\lambda = - \ln (1 - s)$ is recovered for $\alpha=1$, 
and $\alpha > 1$ ($\alpha < 1$) describes a situation with 
negative (positive) epistasis (Fig.~\ref{epiland}).

The error 
threshold can be computed in the unidirectional approximation 
(no back mutations towards the master sequence), and in the limit
$N \to \infty$, $\mu \to 0$, for $\alpha=1, q < 1$, it has been shown that the 
critical mutation strength
$\mu_c = N^{-1} \ln [1/(1 - q)]$, which is of exactly the same form
as the sharp peak result (\ref{etsplc}). For $q=1$, a similar analysis 
shows that (\ref{Wiehe1}) displays
an error threshold only when $\alpha < 1$, with a critical mutation
strength given by $\mu_c = N^{\alpha - 1} \lambda$ \cite{Wiehe97}. Note that 
in this case, the correct scaling is obtained in the limit 
$N \to \infty$, $\mu \to 0$ keeping $\mu N^{1-\alpha}$ fixed.

The above results can be understood using the following result for general 
\textit{bounded} Wrightian fitness landscapes
with $0 < W_{\mathrm{min}} \leq W(\sigma) \leq W_{\mathrm{max}} < \infty$. 
For such landscapes, the master sequence is lost from the population 
at a critical
mutation probability which satisfies (in the unidirectional approximation
and for $N \to \infty$) \cite{Wiehe97}
\be
\l{mucbound}
\mu_c \leq \frac{1}{N} \ln(W_\mathrm{max}/W_\mathrm{min});   
\ee
a similar result is proved in \cite{Wagner93}. If the right hand side of 
(\ref{mucbound}) diverges as $N \to \infty$, this would imply that there 
is no finite error threshold and the master sequence is maintained at
any mutation rate while its vanishing would be consistent with 
the existence of a sharp transition for $\mu \to 0$, $N \to \infty$.
For $q=1$, the ratio between the
largest and the smallest fitness is  
$W_\mathrm{max}/W_\mathrm{min} = e^{\lambda N^\alpha}$, 
so that the right hand side of (\ref{mucbound})
vanishes for $N \to \infty$ only when $\alpha < 1$ 
whereas it goes to zero for any $\alpha > 0$ for $q < 1$, 
in agreement with the results cited above. The case of the multiplicative 
landscape (\ref{Fuji}) is special; here 
$W_\mathrm{max}/W_\mathrm{min} = e^{2 \lambda N}$ and (\ref{mucbound}) would 
suggest 
a finite error threshold\footnote{The unidirectional
approximation erroneously predicts a transition at
$\mu_c = s$ \cite{Wiehe97}.}. However, as discussed earlier, 
the master sequence is maintained at
any mutation rate for $\alpha=q=1$. 

The general conclusion from these considerations is that the 
existence of an error threshold requires positive epistasis. 
This can be understand from the following qualitative argument 
\cite{Wiehe97}: For the case of positive epistasis, the selection 
force towards the fitness peak that has to be overcome by mutations
is largest close to the peak; once this initial barrier has been surpassed,
the population delocalises completely. In contrast, for negative 
epistasis, each additional step away from the fitness peak requires a larger
mutation pressure than the previous step, and hence the population remains
localised.     

\subsubsection{Epistasis: Paramuse models}

Since Malthusian fitness is essentially the logarithm of  
Wrightian fitness, 
the absence of epistatic interactions in continuous time models 
implies a \textit{linear} dependence of the fitness 
$w(\sigma)$ on $d(\sigma,\sigma_0)$. To investigate the 
effects of epistasis, a quadratic fitness landscape of the
form 
\be
\l{quad}
w(\sigma) = a [ 1 - 2d(\sigma,\sigma_0)/N ]
+ \frac{1}{2} b [1 - 2d(\sigma,\sigma_0)/N ]^2 
\ee
has been considered \cite{Baake01}, with $a > 0$ and $b > 0$ ($b < 0$)
for positive (negative) epistasis. This choice of parameters leads,
through the mapping described in Sect.~\ref{muse}, to the quantum 
spin Hamiltonian
\be
\l{Qspin}
H = \tilde \mu \sum_{i=1}^N (\sigma_i^x - 1) + a \sum_{i=1}^N 
\sigma_i^z + \frac{b}{2N} \sum_{ij} \sigma_i^z \sigma_j^z,
\ee
where $\sigma_i^x$ and $\sigma_i^z$ denote the $x$- and $z$-components
of the quantum mechanical spin operator. As in the discrete time case,
in the absence of epistasis ($b=0$) the spins at different sites $i$
are independent. Epistasis introduces a coupling between any pair 
$i,j$ of spins, independent of their position in the sequence. In the
language of statistical mechanics, this is an interaction of 
\textit{mean field} type; it is ferromagnetic for $b < 0$ and 
antiferromagnetic for $b>0$.

An explicit solution of the model has been presented for the case
$a = 0, \; b > 0$ \cite{Baake01}. 
In the limit $N \to \infty$, the mean overlap 
(\ref{overlap}) is given by the expression
\be
\l{overlapBaake}
m = \max[1 - \tilde \mu/b, 0],
\ee
which, in contrast to the case of the sharp
peak landscape,  vanishes \textit{continuously} at $\tilde \mu = b$.
In general, an error threshold exists only if
$- b \leq a < 0$. This implies that the fitness displays a minimum
at a distance $0 < d_{\mathrm{min}} \leq N/2$ from the master sequence.

\subsection{Beyond the standard model}
\l{extensions}

In this section we discuss
a few biologically motivated generalisations
of the mutation-selection models described so far, while however maintaining
the basic simplicity of the fitness landscape.

\subsubsection{Diploid models}

The evolution equations 
for diploid organisms are similar to 
those for the haploid case, except that the fitness $W(\sigma)$ 
is replaced by the marginal fitness 
\be
\l{diploidfit}
\tilde W(\sigma,t)=\sum_{\sigma{'}} W(\sigma,\sigma{'}) X(\sigma',t),
\ee
where 
$W(\sigma,\sigma{'})$ is the fitness of an individual with diploid
genotype $(\sigma, \sigma{'})$, and $X(\sigma,t)$ is the fraction
of individuals carrying sequence $\sigma$ in either one of their
two sets of genes \cite{Wiehe95,Baake97a}. 
The analog of the sharp peak landscape (\ref{spl}) is given by 
\be
W(\sigma,\sigma{'})= \cases {W_{0} & {$: \sigma=\sigma{'}=\sigma_{0}$} \cr
W_{1} & {$:  \mbox{either} \;\sigma = 
\sigma_0 \; \mbox{or} \;\sigma{'}= \sigma_{0}$} \cr
W_{2} & {$: \mbox{both} \; \sigma,\sigma{'} \neq \sigma_{0}$}}
\ee
with $W_0 \geq W_1 \geq W_2$. In the absence of dominance effects
($W_1 = \sqrt{W_0 W_2}$ for Wrightian fitness or
$w_1 = (w_0 + w_2)/2$ for Malthusian fitness) 
the problem can be reduced to the haploid case. 
However, in general, a transformation to a linear equation, as 
described in Sect.~\ref{muse}, is unknown for the 
diploid case; the equations are inherently nonlinear because of the
dependence of the marginal fitness (\ref{diploidfit}) on the
population distribution. As a consequence, 
there are multiple solutions for the fraction $X(\sigma_{0})$
of wild type individuals. 
Nevertheless, error threshold phenomena occur whose locations 
depend on the relative values of $W_{0}$, $W_{1}$ and $W_{2}$. 
For instance, the critical mutation rate is roughly doubled as compared to 
the haploid model in the case of complete dominance of the 
wild type ($W_{0}=W_{1} > W_{2}$). 

\subsubsection{Semiconservative replication}

While the quasispecies model described in Sect.~\ref{muse} is appropriate for 
organisms with RNA as genetic material, it needs to be amended for 
DNA-based organisms. The genotype corresponding to a 
double stranded DNA molecule can be 
represented by $\{ \sigma, \overline{\sigma} \}$ where $\overline{\sigma}$ 
is the complementary strand of $\sigma$. The replication process involves 
splitting the DNA and pairing each strand with the complementary bases to 
produce two daughter DNA's. Thus, only one strand of the original DNA is
conserved in the daughter DNA. However, copying errors and 
subsequent (imperfect) repair result in a different DNA genotype 
$\{ \sigma{'}, \overline{\sigma}{'} \}$. Thus, the (unnormalised) 
number of individuals of genotype $\{ \sigma, \overline{\sigma} \}$
evolves in time as \cite{Tannenbaum04}
$$
\dot{Z}(\{ \sigma, \overline{\sigma} \},t) =  -  
W(\{\sigma, {\overline{\sigma}} \}) Z(\{ \sigma, \overline{\sigma} \},t)
$$
\be
+ \sum_{\{\sigma{'}, {\overline{\sigma}}'\}} 
(p(\sigma{'} \rightarrow \{\sigma, \overline{\sigma} \})+
p({\overline{\sigma}}{'} \rightarrow \{\sigma, \overline{\sigma}\})) \;
W(\{\sigma{'}, {\overline{\sigma}}{'}\}) 
Z(\{ \sigma{'}, {\overline{\sigma}}{'} \},t)  
\ee
where $p(\sigma{'} \rightarrow \{\sigma, \overline{\sigma} \})$ is 
the probability that parent strand $\sigma{'}$ produces 
$\{\sigma, \overline{\sigma} \}$ and the first term represents the loss of 
the original genome.
For the sharp peak landscape, the error threshold occurs at 
\be
\mu_{c}= \frac{2}{N} \ln \left( \frac{2 W_{0}}{1+W_{0}} \right) \l{etsplsc}
\ee
which saturates for $W_{0} \rightarrow \infty$ unlike (\ref{etsplc}), so that 
the loss of the master sequence can not be avoided by increasing its
selective advantage. This can be traced back to the destruction of the parent
genome in the semiconservative case, which implies that, at sufficiently high
mutation probability per genome, 
increasing the reproduction rate of the master sequence actually accelerates 
its 
extinction.

\subsubsection{Dynamic landscapes}

The assumption of a static fitness landscape is good when 
evolution occurs on short time scales or in long-term, controlled 
experiments in the laboratory. 
However, natural populations are usually subjected to dynamic environments 
such 
as that of pathogens living in a host with a dynamic immune system.
For the problem of formation of quasispecies in the presence of a dynamic sharp
peak landscape, two cases need to be distinguished -- one when the fitness 
$W_{0}$ of the  master sequence $\sigma_{0}$ is fixed but its location shifts
at periodic time intervals of length $\tau$ to a nearest neighbor 
\cite{Nilsson00}, 
and the other in which the location is kept fixed but the height of the peak 
changes 
with time \cite{Wilke01,Nilsson01}.

In the former case, besides the usual upper limit on the mutation rate, an 
analytical approximation of the model shows the existence of a lower limit 
also  \cite{Nilsson00}. The latter arises because when the peak shift occurs, 
at least one individual should be present at the new location so that it can 
replicate and form the quasispecies. For too low mutation rates, this may not 
happen and this effect is likely to be more pronounced for finite populations. 

In the case of a time-dependent peak height $W_{0}(t)$ of the master sequence,
the characteristic time scale $\tau$ of variation of the fitness landscape
must be compared to the response time of the population, which is the inverse
of the relative growth rate of the master sequence compared to its mutants.
When $\tau$ is large compared to the response time the population fraction
at the master sequence follows the landscape quasistatically. For
rapidly changing landscapes the time-averaged population undergoes an error 
threshold
transition at the mutation strength $\mu_c$ given by \cite{Wilke01,Nilsson01}
\be
(1-\mu_{c})^{N}=\left( \frac{\int_{0}^{T} W_{0}(t) dt}{T} \right)^{-1}
\ee
which generalises (\ref{etsplc}) by replacing the static fitness by an average 
over a time interval of length 
$T \gg \tau$. For periodic $W_0(t)$ with period $\tau$  
the fraction $X(\sigma_{0},t)$ also changes 
periodically with the same period but with a phase shift that
increases with decreasing $\tau$ \cite{Nilsson01}. Due to this time lag, the 
master sequence achieves maximum population when its fitness has already 
dropped from 
the maximum amplitude. 

\subsubsection{Parental effects}

Digital organisms are computer programs with 
a set of instructions (genome) including copy commands due to which they can 
be replicated. During the copying process, some instructions can get deleted, 
repeated or replaced. An evolved program can perform complex logic operations 
by using a simple logic operator available to it. Such complex organisms are 
selected by allotting them more CPU time thus increasing their replication 
rate defined as the ratio of the number of logical instructions that 
they can execute to the number of instructions that they have to perform
in order to produce a new program \cite{Wilke02c}. While the latter depends 
on the individual's own genome, the CPU time available to it is a parental 
influence. 

The situation is analogous to the case of biological organisms 
which obtain proteins etc. from the parent besides the genome.
In such a case, 
the  fraction $X(\sigma{'},\sigma,t)$ of population at sequence $\sigma$ with 
ancestor $\sigma{'}$ evolves as \cite{Wilke02a}
\be
\l{parental}
\dot{X}(\sigma{'},\sigma,t)=\sum_{\sigma{''}}  A(\sigma{''}) W(\sigma{'}) 
p(\sigma{'} \ra \sigma) X(\sigma{''},\sigma{'},t)-f(t) X(\sigma{'},\sigma,t)
\ee
where  $f(t)=\sum_{\sigma{''},\sigma{'}}  A(\sigma{''}) W(\sigma{'})  
X(\sigma{''},\sigma{'},t)$ and 
$A(\sigma{'})$ is the contribution to the fitness from the ancestor. 
In the absence of parental effects, 
$A(\sigma)=1$ for all sequences and the original equation (\ref{exc}) is 
obtained for 
$X(\sigma,t)=\sum_{\sigma{'}} X(\sigma{'},\sigma,t)$. This can be generalised 
by weighting the 
population variable by the parental contribution and defining the normalised 
variable
 \be
 X(\sigma,t)=\sum_{\sigma{'}} 
A(\sigma{'}) X(\sigma{'},\sigma,t)/\sum_{\sigma{'},\sigma{''}} 
A(\sigma{'}) X(\sigma{'},\sigma{''},t)
 \ee
which reduces the $\ell^{2 N}$ variables in (\ref{parental})  to $\ell^{N}$. 
Interestingly, in the steady state the 
population $X(\sigma,t)$ obeys the quasispecies equation (\ref{exc}) 
with fitness $A(\sigma) W(\sigma)$. 
Thus the available results for the standard quasispecies model can be 
directly applied to this case. 
In particular, for the sharp peak landscape 
the fraction $X(\sigma_{0},\sigma_{0})$ 
at the master sequence increases 
(relative to the null case when there are no parental effects) 
if the ancestral fitness $A(\sigma_{0}) > A(\sigma)$ for 
$\sigma \neq \sigma_{0}$. It is also 
possible to obtain the opposite trend if the ancestral 
effect is deleterious and has to be   
compensated by the fitness of the individual itself, such as when 
$A(\sigma_{0})< A(\sigma)$ and $W(\sigma_{0}) > W(\sigma)$.

\subsubsection{Heterogeneous mutations}

The accuracy of replication 
depends on enzymes called polymerases which can be present in different 
types with their respective accuracies.
For example, as discussed by 
E. L\'azaro elsewhere in this book,
RNA virus strains that show resistance to certain mutagens
may possess polymerases with a particularly high copying 
fidelity.
In the presence of $p$ polymerases with concentrations $c_{k}$ and replication 
error $\mu_{k}$, $k = 1,...,p$, the mutation probability (\ref{mutprob1}) 
generalises to
\be
p(\sigma' \rightarrow \sigma)= \sum_{k=1}^{p} c_{k} 
\left( \frac{\mu_{k}}{(\ell-1)(1-\mu_{k})}\right)^{d(\sigma,\sigma')} 
(1-\mu_{k})^{N}. \l{mutprob3}
\ee
One may expect that by increasing the concentration of the polymerase with 
low error rate, the error threshold can be increased (even to infinity).  
That this indeed is the case was demonstrated in \cite{Aoki03} for $p=2$ with 
concentration $c$ of an error-free polymerase with replication error 
probability
$\mu_{1}=0$ and $1-c$ of an error-prone polymerase with $\mu_{2}=\mu > 0$. 
For the sharp peak landscape, one can find the fraction 
$X(\sigma_{0})= W_{0} p(\sigma_{0} \ra \sigma_0)/(W_{0}-1)$ 
of the master sequence by neglecting the back mutations as before where 
$p(\sigma_{0} \ra \sigma_{0}) \approx c+ (1-c) e^{-\mu N}$ for
$\mu \to 0$ and $N \to \infty$. Then the master 
sequence can localise the population if 
\be
\mu > \mu_{c}= \frac{1}{N} \ln \left( \frac{1-c}{W_{0}^{-1}-c} \right)
\ee
which reduces to (\ref{etsplc}) for $c=0$ as expected and increases with 
increasing $c$. 
Since the argument of the logarithm should be positive for real 
$\mu$, it follows that $c < c'=1/W_{0}$ and on exceeding $c'$, the master 
sequence continues to localise population for any mutation rate. 


\section{Complex fitness landscapes}
\label{Complex}

\begin{figure}
\begin{center}
\includegraphics[width=10cm,angle=0]{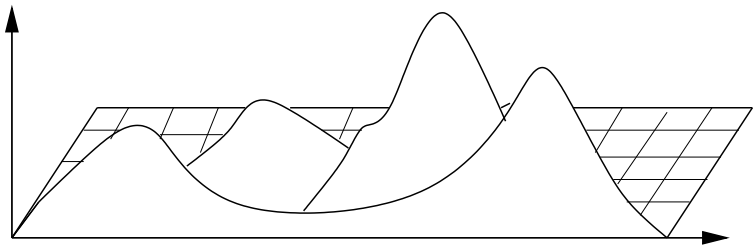}
\caption{Schematic representation of a rugged fitness landscape defined over
a two-dimensional genotype space.}
\label{landscape}
\end{center}
\end{figure}

We now turn our attention to ``complex" landscapes which do not possess 
the symmetries of the simple ones discussed in the last section. Realistic 
landscapes are expected to have hills, valleys, basins and ridges 
\cite{Stadler02}. A pictorial representation of such a \textit{rugged}
fitness landscape drawn over a two-dimensional plane is shown in 
Fig.~\ref{landscape}. Despite the intuitive appeal of such pictures, however,
it should be kept in mind that they are metaphors rather than models
of biological reality. Real fitness landscapes 
extend over the very high dimensional, discrete space of 
genotype sequences, and there are indications that  
the intuition gained in our experience with low-dimensional landscapes
fails when applied to such abstract objects \cite{Gavrilets04}. 

Researchers trying to construct realistic fitness landscapes
have followed one of two basic approaches. One approach is to study
simple model systems for which the mapping from genotype to phenotype can
be carried out explicitly. This has been pursued in great detail for the
case of RNA sequences, which will be briefly described in 
Sect.~\ref{RNAseq}, 
as well as for proteins;
for a detailed discussion we refer to the
chapters by P. Schuster and P. Stadler, and by U. Bastolla, M. Porto,
H. E. Roman and M. Vendruscolo in this book.  
The second approach, which was conceptually 
inspired by the statistical physics of disordered
systems \cite{Anderson83,McCaskill84}, is to regard a given fitness landscape
as the realization of an \textit{ensemble of random functions}
with prescribed statistical properties. In this case an important
quantity characterising the ruggedness of the landscape is
the correlation coefficient $\rho(d,N)$ between the fitnesses of two
genotypes at Hamming distance $d$, which is   
defined as 
\be
\rho(d,N)= \frac{ \langle w(\sigma) w(\sigma') \rangle- \langle w(\sigma) 
\rangle^{2}}{\langle w(\sigma)^2 \rangle- \langle w(\sigma) \rangle^{2}},
\;\; d = d(\sigma,\sigma').
\l{correlation} 
\ee
Here the angular brackets stand for an average over the ensemble of 
landscape configurations and the denominator ensures that $\rho(0,N)$ is 
scaled to unity.
We have defined (\ref{correlation}) in terms of Malthusian fitness, but 
the Wrightian case can be treated in the same way.
Examples of random fitness landscapes will be discussed in Sections 
\ref{uncorrelated}, \ref{correlated} and \ref{neutral}.  

\subsection{An explicit genotype-phenotype map for RNA sequences}
\l{RNAseq}

For the description of evolution experiments with self-replicating RNA 
mole-cules 
(see Sect.~\ref{invitro}), it is natural to assume that the fitness of a given 
RNA 
sequence depends only on the three-dimensional shape that the molecule
folds into in the solution. As an approximation to the full three-dimensional
shape (the \textit{ternary} structure of the molecule), its 
\textit{secondary} structure, defined 
as the set of allowed base pairings that satisfies the no-knot 
constraint and minimises the free energy, can be used. 
In contrast to ternary structure, the secondary structure can be 
computed from the sequence by efficient algorithms.  
Although this does not yet solve the problem of how to assign
a fitness to the genotype, it allows to study in great detail the 
mapping from the genotype (the sequence) to the phenotype
(the secondary structure) \cite{Schuster02,Fontana93,Schuster94}. 

The most important feature of this mapping is that it is 
\textit{many-to-one}. Indeed, the number of secondary structures
of random RNA sequences of length $N$ behaves asymptotically as
\cite{Schuster94}
\be
\l{RNAnumber}
{\cal{N}}_{\mathrm{RNA}} \approx 1.4848 \times N^{-3/2} \times
(1.8488)^N,
\ee
whereas the number of sequences is $4^N$. Thus exponentially many sequences
fold into the same secondary structure for large $N$. Since sequences
with the same secondary structure must be assigned the same fitness, 
it follows that the fitness landscape contains large regions of constant
fitness, which are therefore selectively \textit{neutral}. 
Typically there are a few common structures (which are represented by many
sequences) and many more rare ones, with the distribution of the number
of sequences mapping to a given structure following a power law. The most
common structures form \textit{neutral networks} extending throughout sequence
space, such that any randomly chosen sequence is close to a sequence on this
network. 
Similar networks have also been found in the sequence space
of proteins \cite{Bastolla99,Bastolla03a,Bastolla03b}, 
see the chapter by U. Bastolla, M. Porto, H.E. Roman and M. Vendruscolo
in this book. 
Some aspects of the evolutionary process on such neutral networks
will be discussed below in Sections \ref{neutral} and \ref{Peakshifts}.
  

\subsection{Uncorrelated random landscapes}
\l{uncorrelated}

The simplest kind of random fitness landscape is the uncorrelated 
landscape where the fitnesses are independent random variables drawn
from some common probability distribution \cite{McCaskill84}. 
In this case the correlation
function (\ref{correlation}) reduces to  $\rho(d,N) = \delta_{d,0}$.
An example from this class is the Random Energy Model (REM)
of spin glass theory \cite{Derrida81,Amitrano89,Franz93}, for which 
the (Wrightian) fitness is given by 
\be
\l{kappa}
W(\sigma) = \exp[\kappa E(\sigma)],
\ee
where the ``energies'' $E$ are independent Gaussian random variables with 
distribution 
\be
\l{Gauss}
P(E) = \frac{1}{\sqrt{\pi N}} \exp(-E^2/N),
\ee
and $\kappa$ is an ``inverse selective temperature''. 

This model displays a phase transition
which is quite similar to the error threshold in the single peak landscape.
At high mutation rates the population is delocalised while at low 
mutation rates it is frozen into the master sequence, which in this case
is simply the sequence $\sigma_{\mathrm{max}}$ with the largest value 
$E_{\mathrm{max}}$ of $E(\sigma)$ in the
particular realization (the ``ground state'' configuration of the REM).
The scaling with $N$ in (\ref{Gauss}) 
is chosen such that this maximal value is proportional to $N$,
$E_{\mathrm{max}} =  N \sqrt{\ln 2}$ to leading order.
At the transition the mean overlap (\ref{overlap}) jumps discontinuously from
one to zero \cite{Franz97}. 

The critical mutation probability required for delocalisation can be 
computed along the lines used in Sect.~\ref{Sharppeaksection} 
for the sharp peak landscape.
Neglecting back mutations to $\sigma_{\mathrm{max}}$, a nonzero population
fraction $X(\sigma_{\mathrm{max}})$ is maintained if the product of 
$W_{\mathrm{max}} = \exp[\kappa E_{\mathrm{max}}]$ with the probability
$(1 - \mu)^N$ of producing an error-free offspring is greater than the
mean population fitness $\bar W$ in the delocalised phase \cite{Drossel01}.
The latter is obtained by averaging (\ref{kappa}) with respect to the
distribution (\ref{Gauss}), which yields $\bar W = \exp[\kappa^2 N/4]$.
Comparing the two expressions, one finds \cite{Drossel01,Franz97,Franz93}
\be
\l{mucREM}
\mu_c = 1 - \exp[\kappa^2/4 - \kappa \sqrt{\ln 2}].
\ee
The critical mutation probability reaches its maximal value $\mu_c = 1/2$
at the value $\kappa_c = 2 \sqrt{\ln 2}$ of the inverse selective temperature,
which coincides with the glass transition of the REM \cite{Derrida81}.
For $\kappa > \kappa_c$ the selective advantage of
the most fit sequence is so great that it dominates the population
even in the limiting case $\mu = 1/2$, when a complete reshuffling
of genotypes occurs in each generation.  

We note that, in contrast to most examples discussed in
Sect.~\ref{Simple}, the expression (\ref{mucREM}) is independent of 
the sequence length $N$. This is a consequence of the scaling of
the random energies in (\ref{Gauss}). Indeed, this scaling implies
that the ratio $W_\mathrm{max}/W_\mathrm{min} = \exp[2 E_\mathrm{max}]$
grows exponentially in $N$, and hence the right hand side of 
(\ref{mucbound}) is independent of $N$.

\subsection{Correlated landscapes}
\l{correlated}
 
An example of a random fitness 
landscape with correlations can be constructed from the 
Sherrington-Kirkpatrick (SK) spin glass model, which 
is defined by the energy function 
\be
E_{\mathrm{SK}}(\sigma)=\frac{1}{N} \sum_{i < j} J_{ij} \sigma_{i} \sigma_{j}.
\l{SK}
\ee
Here $\sigma_{i}=\pm 1$ and the $J_{ij}$ are independent Gaussian  
random variables with 
zero mean and unit variance. 
A similar energy function arises for the graph bipartitioning problem (GBP)
discussed in Sect.~\ref{basic}, 
\be
E_{\mathrm{GBP}}(\sigma)=-\sum_{i < j} J_{ij} \sigma_{i} \sigma_{j},
\l{GBP}
\ee
where the spins satisfy the vanishing total spin constraint. In this case
$J_{ij} = J > 0$ if the sites $i$ and $j$ are connected by an edge of the 
graph,
and $J_{ij} = 0$ else \cite{Fu86,Bonhoeffer93}. Through (\ref{kappa}) 
energy functions (\ref{SK}) and (\ref{GBP})
can be directly interpreted as Malthusian fitness
landscapes \cite{Anderson83,Amitrano89,Bonhoeffer93}.
They belong to a large class of random landscapes for which the correlation 
function
behaves as \cite{Fontana93} 
\be
\rho(d,N)
\approx 1- a_{1} \frac{d}{N}  +{\cal{O}} 
\left( \left( \frac{d}{N}\right)^{2} \right) \l{corr}
\ee
for $N, d \to \infty$ but $d/N \ll 1$,
with a constant $a_{1}$ which is independent of $N$. 
The significance of this behavior becomes clear if we interpret $d/N$ as a 
continuous variable: For random functions of a real variable,
the linear dependence of the correlation function for small arguments is 
typical
of a non-differentiable process with independent increments (such as Brownian 
motion), 
whereas for a differentiable random process the correlation function 
varies quadratically at small distances. In this sense
the linear behavior in (\ref{corr}) is indicative of the ruggedness of the
landscape.   

A simple modification of the argument leading to (\ref{mucREM}) gives some
insight into how the fitness correlations affect the location of the 
error threshold \cite{Bonhoeffer93}. We assume that in the localised phase
the bulk of the population is located at some distance $d^\ast = 
{\cal{O}}(1)$ from the most fit genotype $\sigma_{\mathrm{max}}$, with
corresponding energy values $\bar E \approx \rho(d^\ast) E_\mathrm{max}$.
Equating the resulting mean population fitness $\bar W = \exp[\kappa 
\bar E]$ to the product $(1 - \mu)^N W_\mathrm{max}$ and using
(\ref{corr}) then yields, for large $N$, the estimate
\be
\l{muccorr}
\mu_c \approx \frac{\kappa a_1 d^\ast E_\mathrm{max}}{N^2}.
\ee
Together with the scaling of the ground state energy 
as $E^{\mathrm{SK}}_{\mathrm{max}} \sim N^{1/2}$ and 
$E^{\mathrm{GBP}}_\mathrm{max} \sim N^{3/2}$ for the SK-model
and the GBP, it follows that $\mu_c^{\mathrm{SK}} \sim N^{-3/2}$
and $\mu_c^{\mathrm{GBP}} \sim N^{-1/2}$, respectively, in agreement with
simulations \cite{Bonhoeffer93}.

\begin{figure}
\begin{center}
\includegraphics[width=9.cm,angle=0]{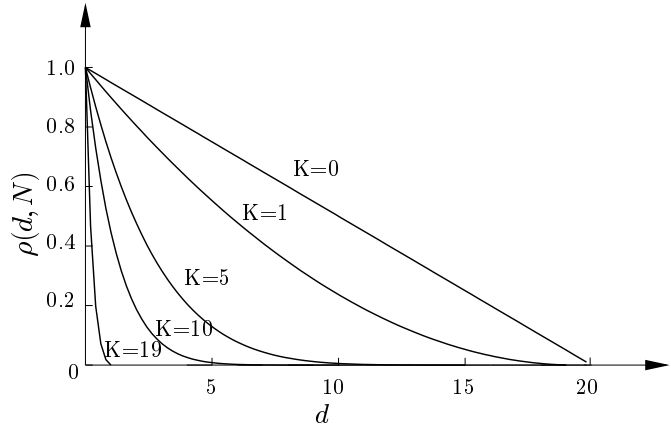}
\caption{Correlation function (\ref{NKcorr}) for the $NK$-model with $N = 20$.}
\label{NKmodelfig}
\end{center}
\end{figure}

A family of random landscapes in which the ruggedness can be tuned
are the $NK$ landscapes\footnote{A related family was defined in 
\cite{Amitrano89} in analogy to Derrida's $p$-spin model of 
spin glasses \cite{Derrida81}.} 
introduced by Kauffman and Levin \cite{Kauffman87,Kauffman93}. 
In this model, the Malthusian fitness\footnote{A Wrightian version of the 
model is 
discussed in \cite{Welch05}.} 
of a genotype is written as a sum of contributions from the $N$ loci,
\be
\l{NKmodel}
w(\sigma) = \frac{1}{N} \sum_{i=1}^N w_i,
\ee
where each $w_i$ is a function of $\sigma_i$ and $K$ 
other loci chosen at random\footnote{Other schemes for choosing the 
interacting loci 
are described in \cite{Perelson95,Fontana93,Welch05}.}.
The number of possible states of $\sigma_i$ and its $K$ chosen neighbors is 
then
$\ell^{K+1}$, and each of these states is assigned a random fitness drawn 
from some
continuous probability distribution. For large $N$ the additive form of 
(\ref{NKmodel}) ensures that the $w(\sigma)$ become Gaussian by virtue of the
central limit theorem.

For $K=0$ the loci are independent, and the model becomes equivalent to the 
multiplicative fitness landscape without epistasis discussed in 
Sect.~\ref{Modified};
in particular, there is a unique fitness peak. At the other extreme
$K = N-1$, the $w_i$ are independent random variables and the model reduces
to the uncorrelated landscape of Sect.~\ref{uncorrelated}. With increasing
$K$ the number of fitness maxima increases and their height decreases
\cite{Kauffman93}, and the 
correlation function is given by\footnote{The expressions for the correlation 
function given in \cite{Fontana93,Weinberger91} are incorrect, because it is 
not taken into
account that the $d$ mutations separating the two genotypes in 
(\ref{correlation})
must affect different sites in the sequence.} 
\cite{Campos02}
\be
\l{NKcorr}
\rho(d,N) =   \left\{ 
\begin{array}{ll}
\frac{(N - K - 1)! \; (N-d)!}{N! \;
(N - K - d - 1)!} & : d \leq N - K - 1 \\
0 & : \mathrm{else}. 
\end{array} 
\right.
\ee
This shows how the correlations decay more rapidly
with increasing epistasis (increasing $K$),
and reduces to $\rho(d,N) = 1 - d/N$ for $K=0$ and $\rho(d,N) = \delta_{d,0}$ 
for
$K = N-1$, respectively (see Fig.~\ref{NKmodelfig}).

Another model with tunable correlations was introduced in a study of 
evolutionary 
dynamics in the limit of infinite genome size but with a finite population 
\cite{Wilke02b}.
For $N \to \infty$ every mutation creates a genotype that has not been 
previously represented in the population. The fitnesses can then 
be created ``on the fly'' according to the transition probability
\be
\mathrm{Prob}[w(\sigma)|w(\sigma')] \sim 
\mbox{exp}[-(w(\sigma)-\lambda^{d} w(\sigma'))^{2}]
\ee
where $d = d(\sigma,\sigma')$ and
the parameter $0 \leq \lambda \leq 1$ determines the decay of the correlations
as $\rho(d,N) \sim \lambda^d$.


\subsection{Neutrality}

\l{neutral}

We have seen above in Sect.~\ref{RNAseq} that realistic fitness landscapes
obtained from mapping sequences to structures contain extended regions
that are selectively neutral. It has been argued that this is a general feature
of high-dimensional fitness landscapes, which has important consequences for
the way in which evolutionary dynamics should be visualized \cite{Gavrilets04}.
Rather than consisting of valleys and hilltops, as suggested by the 
low-dimensional
rendition in Fig.~\ref{landscape},
such a \textit{holey landscape} 
would display a network of ridges of approximately constant
fitness, along which a population can travel large genetic distances without
ever having to cross an unfavorable 
low-fitness region\footnote{The evolutionary importance of paths of viable 
genotypes that connect distant points in sequence space 
was emphasized by Maynard Smith \cite{Smith70}. He illustrates the issue with
a game where the goal is to transform one word into another by changing
one letter at a time, with the requirement that all intermediate words 
are meaningful (i.e., ``viable''). An example is the path WORD $ \rightarrow$
WORE $ \rightarrow$ GORE $ \rightarrow$ GONE
$ \rightarrow$ GENE.}.

Several properties of the stationary population distribution for the 
quasispecies model on a neutral network can be inferred without
specifying the precise structure of the network \cite{vanNimwegen99}. 
It is only assumed that
the viable genotypes make up a connected graph 
${\cal{G}}$ of constant fitness, which is
surrounded by genotypes that are lethal or at least of very low fitness.
Mutations are restricted to nearest neighbor sequences. Then  
the key observation is that the stationary population distribution $X(\sigma)$
on the network\footnote{The population on the network is normalized to unity,
$\sum_{\sigma \in {\cal{G}}} X(\sigma) = 1$, which does not include the 
individuals in the lethal region. Although these individuals do not reproduce,
they constitute a finite fraction of the population which is replenished by
mutations from viable genotypes.}  
is the principal eigenvector of the \textit{adjacency matrix} of the graph,
which is a matrix that has unit entries for pairs of viable sequences that 
are connected by a single point mutation, and zero entries otherwise. The 
corresponding
eigenvalue $\Lambda$ is equal to the \textit{population neutrality}
$\langle \nu \rangle$,
\be
\l{neutrality}
\Lambda = 
\langle \nu \rangle = \sum_{\sigma \in {\cal{G}}} \nu(\sigma) X(\sigma),
\ee
where $\nu(\sigma)$ is the number of viable neighbors of sequence $\sigma$
(the \textit{degree} of the corresponding node of ${\cal{G}}$).
The weighting by the population fraction $X(\sigma)$ in (\ref{neutrality})
is significant: For any graph ${\cal{G}}$ the principal eigenvalue of the
adjacency matrix satisfies the bounds \cite{Soshnikov03}
\be
\l{adjbounds}
\bar \nu \leq \Lambda \leq \nu_{\mathrm{max}},
\ee
where $\bar \nu$ and $\nu_{\mathrm{max}}$ denote the average and maximal 
degrees
of the graph. For a random graph with a range of degrees the relations 
(\ref{neutrality}) and (\ref{adjbounds}) imply that generally
$\langle \nu \rangle > \bar \nu$, which shows that the population 
preferentially
resides at nodes where the number of viable neighbors is larger than on 
average.
This has been referred to as the evolution of \textit{mutational robustness}
\cite{vanNimwegen99}.
The heterogeneity of the node degree along the neutral network has
important consequences also for the evolutionary dynamics, 
because it induces strong fluctuations
in the rate of neutral substitutions \cite{Bastolla03a,Bastolla03b,Bastolla02}.

Neutral networks can be modeled as random subgraphs in
sequence space. Such subgraphs are defined through a simple modification 
of the uncorrelated landscape model of Sect.~\ref{uncorrelated}, where
each sequence $\sigma$ is randomly assigned fitness $W(\sigma) = 1$ 
(\textit{viable})
with probability $P$ and $W(\sigma) = 0$ (\textit{lethal}) with 
probability $1 - P$. Each connected region of viable genotypes then constitutes
a random subgraph. For small $P$ these regions are small and isolated, but
at the \textit{percolation threshold} $P = P_c$ given by 
\be
\l{Pc}
P_c = \frac{1}{(\ell - 1)N}
\ee
a giant network appears which spans the sequence space and which, for
$P > P_c$, contains a finite fraction of all sequences 
\cite{Gavrilets04,Gavrilets97}. 
Since $N$ is a large number, the fraction of viable genotypes needed
to create such a spanning network is remarkably small \cite{Smith70}.

For subgraphs of the binary 
hypercube ($\ell = 2$) with random assignment of links (rather
than sites) it has been shown that the principal eigenvalue
of the adjacency matrix is asymptotically given by \cite{Soshnikov03}
\be
\l{numax}
\Lambda \approx \max[N P, \sqrt{\nu_{\mathrm{max}}}].
\ee
Taking $N \to \infty$ at fixed $P$ one finds that $\nu_{\mathrm{max}} \sim N$, 
so that $\Lambda \to N P = \bar \nu$. In this limit the neutral network behaves
like a regular graph, and no significant mutational robustness develops.
On the other hand, if $P \to 0$ as $N \to \infty$ with $NP$ fixed,
one obtains $\nu_{\mathrm{max}} \sim N/\ln N \gg \bar \nu$, and 
the mutational robustness effect is significant.


\section{Dynamics of adaptation}
\label{Dynamics}

In this section we turn our attention to time-dependent aspects of the adaptive
process. In rugged fitness landscapes the
population is faced with the task of reaching ever higher fitness peaks by
traversing fitness valleys or neutral networks, which typically gives rise to 
a pattern of episodic or \textit{punctuated} evolution. This phenomenon will be
discussed in general terms in the following subsection, and a specific model 
study
\cite{Jain05} will be summarized in Sect.~\ref{punctuation}. In the final 
subsection
we describe an approach to evolutionary dynamics that is suited for landscapes
that are \textit{smooth}, in the sense that a simple (linear) relation between
fitness and genetic distance can be assumed. 

\subsection{Peak shifts and punctuated evolution}  
\label{Peakshifts}

The existence of multiple fitness peaks of different height, as illustrated 
in Fig.~\ref{landscape},
immediately suggests that evolutionary histories
should generally display two distinct regimes: Periods of stabilizing 
selection, where the population
resides near a local fitness maximum, and \textit{peak shifts} in which the 
population moves quickly from one
fitness peak to another of greater height. The stationary distributions in a 
single peak landscape that
were discussed at length in Sect.~\ref{Simple} can be viewed as an approximate 
description of the first
regime. The necessity of peak shifts for explaining the succession of 
biological forms in the 
paleontological data has been recognized for a long time\footnote{A famous 
example is the transition
from browsing to grazing behavior in equids \cite{Simpson44}.}, 
but the underlying mechanisms (and even the relevance of the concept itself) 
remain controversial.   

Mathematical analysis of peak shifts driven by stochastic fluctuations in 
finite 
populations (genetic drift) generally show that the waiting time for the 
shift is vastly larger
than the time required for the transition itself \cite{Newman85,Barton87}. 
This can be argued to 
support the scenario of \textit{punctuated equilibrium} in macroevolution 
\cite{Eldredge89}, 
which states that evolutionary 
changes (including both speciation and phenotypic changes within a lineage) 
occur during 
relatively short time intervals which are separated by long periods of no 
discernible change
(\textit{stasis}).

However, for realistic population sizes the stochastically driven peak shifts 
may be far too rare
to be relevant, and in fact they may not be needed at all, if the picture of 
a holey landscape
spanned by a network of neutral ridges described in Sect.~\ref{neutral} is 
generally
applicable \cite{Gavrilets04}. Evolution in such a landscape 
will nevertheless be punctuated, because a population
moving by genetic drift across a neutral network can increase its fitness 
only by finding a path
to another network of higher fitness. If these paths are rare, a natural 
separation of time scales
between (phenotypic) stasis and sudden fitness jumps arises. This scenario is 
well established
for simulations of \textit{in vitro} evolution of RNA sequences 
\cite{Schuster02,Fontana98}.  
Borrowing a concept from statistical physics, it can be said that in this 
case the population
is confined by \textit{entropic} barriers rather than by fitness barriers 
\cite{vanNimwegen00}.


\subsection{Evolutionary trajectories for the quasispecies model}
\l{punctuation}

\begin{figure}
\begin{center}
\includegraphics[width=4.cm,angle=-90]{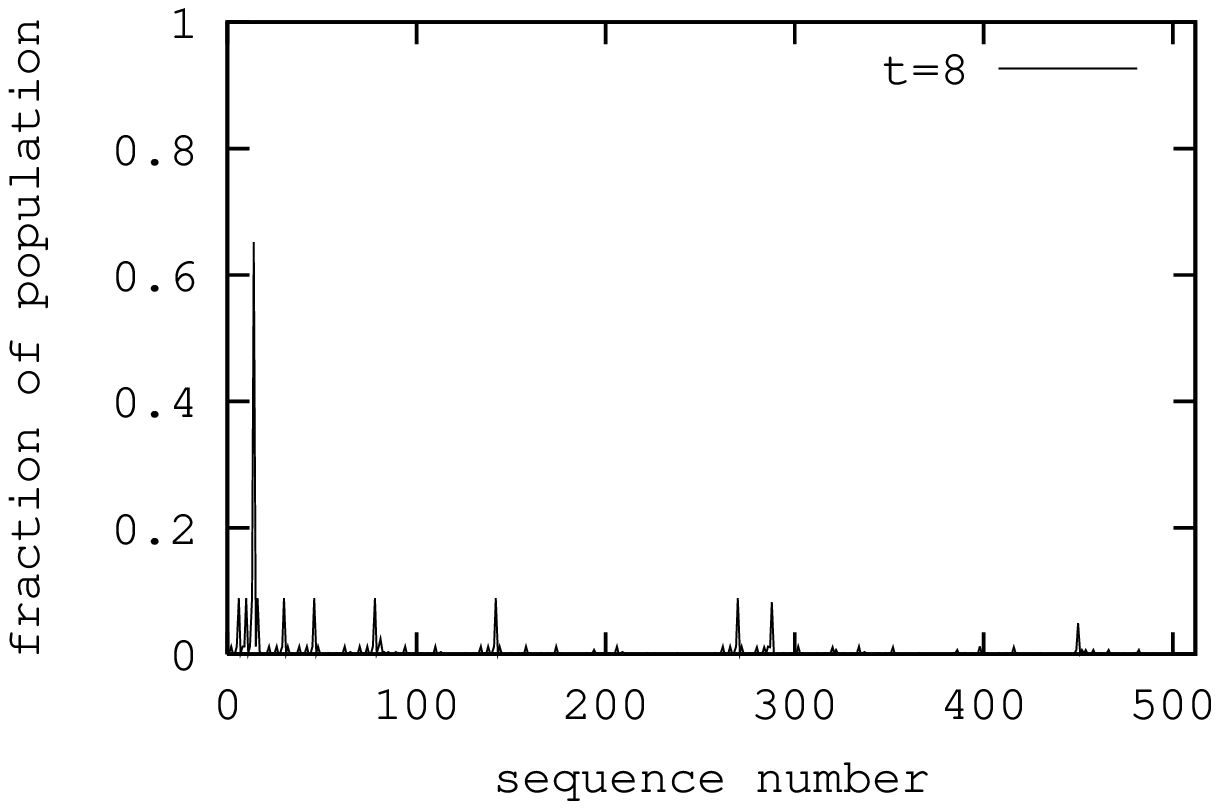} \includegraphics[width=4.cm,angle=-90]{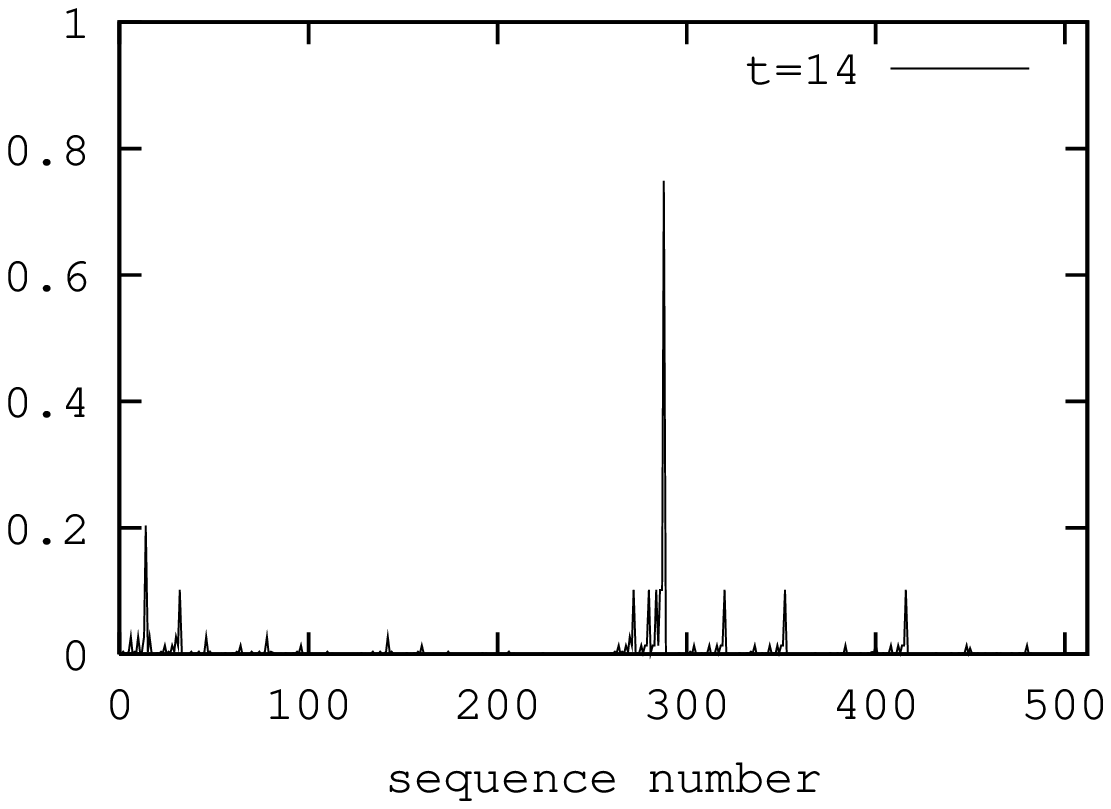}
\caption{Example of a peak shift event for quasispecies dynamics with binary 
sequences of length
 $N=9$ in an uncorrelated random
fitness landscape. At time $t=8$ the most populated sequence is near the 
origin (sequence number
1), but at time $t=14$ it has moved to a sequence number close to 300. The 
peaks of lower height
represent the first- and second neighbor mutants of the most populated 
sequence. They are not
adjacent because of the linear arrangement of the sequences.}
\label{PeakShift}
\end{center}
\end{figure}

In the deterministic mutation-selection models of interest in this chapter, 
stochastic fluctuations
cannot be invoked to drive peak shifts. Nevertheless a population initially 
placed near one fitness peak 
in a multi-peaked landscape is able to relocate to a higher peak, by 
developing tails
of mutants which (since the number of individuals is formally infinite) with 
time explore the 
entire sequence space. Once a small mutant population has been established at 
the distant fitness
peak, it starts to compete with the majority at the original peak and, if the 
newly populated peak
is higher, it will eventually come to dominate the population. In this way 
the majority of the 
population can shift between peaks without ever actually having to traverse a 
fitness valley
(Fig.~\ref{PeakShift}). 

The time $t_\times$
required for a single peak shift in the discrete time quasispecies model has 
been estimated 
numerically for a simple degenerate two-peak landscape, given by 
(\ref{twopeaks}) 
with $W_N = W_0$ and $W_{N-1} = 1$ \cite{Kim04}. The population was first 
allowed to 
equilibrate in a single peak landscape and then the second peak was turned on.
The result is 
\be
\l{peakshifttime}
t_\times \sim \left( \frac{\ln W_0}{N \mu} \right)^N 
\sim \left(\frac{\mu_c}{\mu} \right)^N
\ee 
which, somewhat surprisingly, has the same form as the time required for a 
finite 
population to cross a fitness valley \cite{vanNimwegen00}; of course in the 
latter case
there is an additional dependence on the population size.

The evolutionary trajectories that result from multiple peak shifts in an 
uncorrelated rugged
fitness landscape have been studied in detail in a strong selection limit 
motivated by the
zero temperature limit of the statistical physics of disordered systems 
\cite{Jain05,Krug02,Krug03,Krug05}. 
Writing 
\be
\l{strong}
Z(\sigma, t) = e^{\kappa F(\sigma, t)}, \;\; W(\sigma) = e^{\kappa E(\sigma)},
 \;\; 
\mu=e^{-\kappa} ,
\ee
with $\kappa$ denoting the inverse selective temperature 
(see Sect.~\ref{uncorrelated}),
and starting with an initial condition 
$Z(\sigma,0)=\delta_{\sigma,\sigma^{(0)}}$ where $\sigma^{(0)}$ is a randomly 
chosen sequence, the dynamics takes the following form in the 
$\kappa \ra \infty$ limit:
\bea
F(\sigma,t+1) &=& \mbox{max}_{\sigma'} \left[F(\sigma',t)+
E(\sigma')-d(\sigma,\sigma') \right], \;\; t \ge 2  \\
F(\sigma,1) &=& E(\sigma^{(0)})-d(\sigma,\sigma^{(0)}).
\eea
Here the logarithmic fitnesses $E(\sigma)$ are independent random
variables chosen from a common distribution $p(E)$. As already discussed 
in Sect.~\ref{uncorrelated}, one expects 
the whole population to be localised at the fittest 
genotype in the large time limit. At any finite time, 
in the strong selection limit, the population can 
be identified with the most populated genotype. The behavior
of this genotype is essentially unaffected
by dropping the mutation term for times $t > 1$, so that
the dynamics reduces to \cite{Krug03}
\be
F(\sigma,t)=F(\sigma,1)+(t-1) E(\sigma), \;\; t \geq 2.
\l{linear}
\ee
This illustrates the fact that, after the entire sequence space has been
``seeded'' by mutants of the original genotype $\sigma^{(0)}$
at time $t = 1$, the subsequent
evolution consists in the competition of independent populations located
at the fitness peaks. Distant peaks of high fitness are disadvantaged
by a small initial population but may come to dominate at later times.

Since the seeding population $F(\sigma,1)$ of a sequence only depends on its
distance from the initial genotype $\sigma^{(0)}$, within each \textit{shell}
of constant $k = d(\sigma,\sigma^{(0)})$ only the most fit genotype is a 
contender
for global leadership. Thus the dynamics of the $\ell^N$ variables 
(\ref{linear})
can be reduced to 
$N+1$ shell population 
variables $F(k,t)$ whose fitnesses $E(k)$ are chosen from the distribution 
\be
p_{k}(E)=\alpha_{k} \; p(E) \left( \int_{E_\mathrm{min}}^{E} p(x) dx 
\right)^{\alpha_{k}-1}.
\l{fitmax}
\ee
This is the distribution of the maximum among $\alpha_k$ independent random 
variables
with distribution $p(E)$, and $\alpha_{k}$ is the number of sequences in 
shell $k$, 
as defined in (\ref{alphak}). 

\begin{figure}
\begin{center}
\includegraphics[width=6.cm,angle=-90]{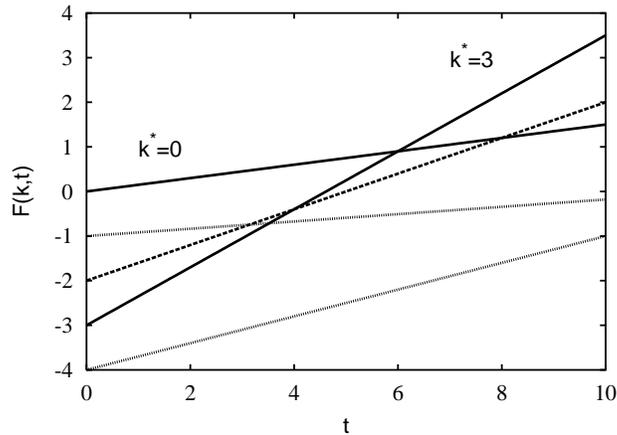} 
\caption{Illustration of the linear dynamics (\ref{linear}). For each shell 
of constant
Hamming distance $k$ from $\sigma^{(0)}$ only the line with the largest slope 
is drawn. Dashed lines
are fitness records, dotted lines are non-records, 
and solid lines are records that are not bypassed.}
\label{Lines}
\end{center}
\end{figure}

The representation of the ``evolutionary race'' as a problem of crossing 
straight lines\footnote{The problem is related to models of highway 
traffic, where each vehicle is equipped with a fixed random speed and 
overtaking is forbidden \cite{BenNaim94}.}
is illustrated in Fig.~\ref{Lines}.
At a given time $t$, the 
most populated sequence located in shell $k^{*}$ leads until it is overtaken 
by a shell $k^{*'}$ with 
$E(k^{*'}) > E(k^{*})$ and so on, until the global fitness maximum takes over.
A natural question of interest is to identify the sequences that take part in 
this
evolutionary trajectory, and to determine their number. It is clear that for
a sequence to participate in the trajectory it is necessary that it constitutes
a \textit{fitness record}, in the sense that its fitness exceeds the fitnesses
of all sequences that are closer to $\sigma^{(0)}$.   
An analytical treatment of the statistics of these independent but 
non-identically distributed records shows 
that the average number of records encountered on the way to the global
maximum is \cite{Jain05}
\be
\l{records}
{\cal{R}} \approx \frac{(\ell - \ln \ell -1)}{\ell - 1} N
\ee
for large $N$, and that essentially all records are located within the 
distance 
$d_{\mathrm{max}} = N (\ell -1) /\ell$ near which most of the sequences 
(including the most fit sequence) reside. For $\ell=2$, the inter-record 
spacing between the $j$-th and $j+1$-th record is of the order 
$\sqrt{N/j}$ where $j=1$ labels the last record (the global maximum). 
Thus, a few records separated by distances of order $\sqrt{N}$ occur near 
$d_{\mathrm{max}}$ and the rest are clustered away from it. 

However, many records are \textit{bypassed} by fitter sequences that arise 
further away from $\sigma^{(0)}$ but manage to catch up with the current 
leader at an earlier
time. For unbounded fitness distributions with Gaussian or exponential tails,
the number of non-bypassed records (which is the number of sequences that take
part in a trajectory) is found to be only of order $\sqrt{N}$ with 
a uniform spacing $\sim \sqrt{N}$, which suggests that the 
competition among the contenders is strong when the average fitness of the 
population is still low. For fat-tailed
power law distributions the average number of records that are not bypassed 
is asymptotically equal to unity, which 
implies
that the population relocates to the global fitness maximum in a single step.

\begin{figure}
\begin{center}
\includegraphics[width=8.cm]{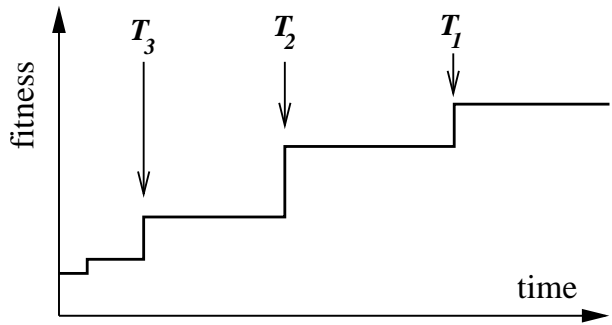} 
\caption{Timing of evolutionary jumps.}
\label{Jumps}
\end{center}
\end{figure}

Several statistical properties of the \textit{timing} of peak shifts turn out 
to be 
independent of the fitness distribution $p(F)$ \cite{Jain05,Krug02,Krug03}. 
Specifically,
denoting by $T_j$ the time at which the $j$th peak shift occurs, with $j = 1$ 
denoting
the last shift (which reaches the global fitness maximum), $j=2$ the 
penultimate peak
shift and so on (Fig.~\ref{Jumps}), the corresponding distributions display 
universal
power law tails
\be
\l{evtimes}
P_j(T_j) \sim (T_j)^{-(j+1)}.
\ee
In particular, the expected value of $T_1$ is infinite. The prefactors of 
these power laws depend
however on the fitness distribution and the sequence length, in such a way 
that e.g.
the typical value of $T_1$ tends to unity for fitness distributions with a 
power law tail.

\subsection{Dynamics in smooth fitness landscapes}
\l{Smooth}

So far we have discussed landscapes in which the fitnesses can be very
different from each other and as described above, the evolutionary
trajectory can change in a stepwise manner if the landscape has local
maxima. Smoothly varying landscapes for which the system does not get trapped
in such metastable states are the subject of the following discussion.
\textit{Smoothness} will be taken to imply here that there is a simple
(linear) relationship between the fitness of a genotype and its
genetic distance from the master sequence. Individuals can then be 
characterized by their fitness alone, and the description can be
based on a one-dimensional \textit{fitness space} \cite{Tsimring96}.

The prototypical case in which this reasoning applies is that of
the multiplicative fitness landscape discussed in Sect.~\ref{Modified}.
We work in the Malthusian setting and assume that the fitness $w(\sigma)$
is simply equal to the number of mismatches with respect to the master
sequence, $w = 0,...,N$.  
Then the fraction $Y(w,t)$ of individuals
with fitness $w$ at time $t$ evolves as \cite{Tsimring96} 
$$
\dot{Y}(w,t)=(w-\overline{w}) Y(w,t)+
$$
\be
+ \tilde \mu [(w+1) Y(w+1,t) + (N - w + 1) 
Y(w-1,t) - N Y(w,t)],
\l{smooth1}
\ee
which is just the paramuse equation (\ref{ckxc}) evaluated for the present 
fitness 
landscape, with $\bar w$ denoting the mean fitness of the population. 
For large $N$ the fitness $w$ can be treated as a continuous variable.
Setting $r = (w - N/2)/\sqrt{N}$, $\bar \mu = \tilde \mu/\sqrt{N}$ and 
$\tau = \sqrt{N} t$, Eq.~(\ref{smooth1}) reduces for $N \to \infty$ to the
drift-diffusion equation  
\be
\l{smooth2}
\frac{\partial Y}{\partial \tau} = (r - \bar r) Y 
+ \frac{\bar \mu}{2} \frac{\partial^2 Y}{\partial r^2}
+ \frac{\partial}{\partial r} ( 2 \bar \mu r Y).
\ee
Analysis of (\ref{smooth1}, \ref{smooth2}) and related equations 
\cite{Kessler97} shows that
the mean fitness diverges
in finite time, since the equations ignore the fact that
at least one individual is required to initiate the reproduction process.
This can be circumvented by imposing a cutoff $Y_{c}$ inversely proportional to
the population size, below
which the selection term does not operate. 

With this modification, one finds that at short times, the
population which was initially spread over a fitness range gets localised
about the maximum available fitness leading to a fast growth of average
fitness. This is followed by the collective motion of the localised
``species'' as a traveling wave with constant speed and width (as long as the 
population is far from the boundaries $w = 0$ and $N$ of the fitness space). 
A finite population size analysis of discrete models (described 
in the infinite population limit by the above continuum equations) shows that 
both speed and variance of the wave diverge linearly with increasing 
population size, which is
consistent with the finite time singularity that appears in the absence of 
a cutoff \cite{Kessler97}. 

Quantitative agreement with finite population simulations requires 
a more careful treatment in which the most fit non-empty mutant class 
is treated stochastically, while keeping deterministic differential-difference
equations of the type (\ref{smooth1}) for the remainder of the population. 
In addition, the continuum limit of (\ref{smooth1}) should be carried out on 
the level of $\ln Y$ rather than for $Y$ itself, which leads to a nonlinear
drift-diffusion equation replacing (\ref{smooth2}) \cite{Rouzine03}. 
Recent applications of fitness space models that go beyond the present 
discussion include studies of the \textit{in vitro} evolution of DNA sequences selected 
for protein binding \cite{Peng03}, 
viral populations undergoing serial
population transfers \cite{Manrubia03},
and the effects of recombination in asexual populations \cite{Cohen05}.


\section{Evolution in the laboratory}
\l{Experiments}

Viruses and bacteria are suitable candidates for testing the theory of asexual 
evolution due to their simple genomes and high replication rates. For 
instance, RNA viruses which are characterised by high mutation rates and 
small genome (see Table \ref{drake} and the chapter by E. L\'azaro
in this book) can produce 
about $10^{4}$ copies an hour. Their typical population numbers are of the 
order of  
$10^{11}$, thus getting close to the infinite population condition for the 
applicability of quasispecies theory. Interestingly, evolution can also 
occur in non-living systems 
such as RNA extracted from a bacteriophage which we now proceed to discuss in 
the 
following subsection. 

\subsection{RNA evolution \textit{in vitro}}
\l{invitro}

Early \textit{in vitro} studies of adaptation to a given environment were carried
out on a simple system comprising of RNA molecules and the enzyme RNA 
replicase which is required to catalyse the RNA replication reaction. In the 
first of a series of experiments, 
the time interval during which the reaction is allowed to 
proceed was gradually reduced with the number of generations, thus 
selecting the rapidly growing molecules \cite{Mills67}. By the 74th 
generation, the initial 
baseline strain with a genome length of a few thousand bases 
evolved to a 15 times faster replicating (but no longer pathogenic) chain 
of merely a few hundred bases,  
by casting off the parts of the genome which do not participate in the
\textit{in vitro} replication process. Subsequently, experiments using such 
short RNA were 
performed under different conditions and selection pressure 
\cite{Biebricher83,Biebricher97}. In particular, the formation of a 
quasispecies consisting only to $40 \%$ of the master sequence and many 
mutants has been demonstrated \cite{Rohde95}.  

\subsection{Quasispecies formation in RNA viruses}
\l{RNAviruses}

Inside a cell, a virus is subjected to the constantly changing environment
of the host, whereas the quasispecies concept described in earlier 
sections assumes an
infinite population evolving towards a stationary state
in a static landscape. Nevertheless, evidence for quasispecies formation has 
been obtained 
in \textit{in vivo} experiments on RNA viruses by examining their genetic 
heterogeneity
\cite{Domingo97}, and the quasispecies concept now plays an important role in 
virology \cite{Nowak01,Eigen02,Moya04};
for a detailed discussion we refer to the
chapter by Ester L\'azaro in this book.   

The first such experiment was performed on a Q$\beta$ phage population derived 
from the wild type \cite{Domingo78}. On sampling about $10 \%$ 
of its sequence,  
it was found that on average, 
the genome of the derived phage differs from the wild type at about two 
positions. Assuming a Poisson model for the distribution of deviations from 
the wild type, only $14 \%$ of the population was found to be wild type and 
the rest was accounted for by related mutants with up to 3-4 substitutions.  
Similarly, in the Hepatitis C virus, half of
the RNA molecules were found to be identical and the rest one to four
mutations away from each other \cite{Martell92}. In the case of HIV, the
quasispecies concept has been used to explain the reappearance of the virus
after the treatment with drugs that target only the wild type
\cite{Coffin95}. Many experiments, such as \cite{Crotty01} on poliovirus,
also show that RNA viruses operate close to the error
threshold, since on a modest increase in mutation rate (through chemicals),
the virus population was found to lose its genetic structure. 

\subsection{Dynamics of microbial evolution}

The dynamics of adaptation have been studied in several 
long-term experiments on asexually reproducing microbes like viruses and 
bacteria. In experiments on \emph{E. coli} \cite{Lenski91,Lenski94}, 
several populations are derived from the same ancestor and allowed to 
replicate under identical conditions. The ancestor is engineered to have 
a selectively neutral marker so that it can be distinguished from the 
offspring colony. The process of evolution occurs because the progeny 
is grown in the presence of limited supply of glucose, unlike the ancestor. 

To measure the fitness of the evolved type, the ancestor 
and the evolved progeny are made to compete for glucose by mixing them in 
equal amounts at time $t=0$ and estimating their respective densities 
$\rho_{A}$ 
and $\rho_{P}$ at $t=0$ and $t=1$ where time is measured in 
days. Then the Malthusian fitness of the evolved type at any instant measured 
relative to the ancestor $A$ is given by 
\be
w = \frac{\ln(\rho_{P}(1)/\rho_{P}(0))}{ \ln(\rho_{A}(1)/\rho_{A}(0))}.
\ee
The experiments indicate that the fitness of all  
populations improves in time, but each of the replicate populations 
reaches a different fitness level at large times. 
This supports the picture of a rugged fitness landscape 
(Fig.~\ref{landscape}) with several peaks 
in which the population, starting from the same initial point, reaches different 
local maxima via different evolutionary trajectories.  

Initially 
fitness changes rapidly but slows down considerably in the course of time. 
When 
the same experimental data is viewed at a finer scale,  
the best fit to the data is obtained if the fitness increases 
are assumed to occur in steps. The occurrence of punctuated evolution is 
associated with the selection of rare beneficial mutations \cite{Elena96}. 
Although a large number of advantageous mutations with small effects may have 
occurred, a few mutations with large effects quickly spread through the 
population 
and are responsible for the jumps in the fitness. For a review of other 
experiments with this bacterial population see \cite{Elena03,Lenski04}.

The step-like nature of fitness trajectories, especially the properties of the
first step, has been investigated in 
detail in other experiments as well. For instance, in \cite{Imhof01}, the 
distribution of the fitness conferred in the first step 
was measured in \emph{E. coli}, which 
supports the above observation of the occurrence of few mutations with 
large benefits and many with small payoffs. Similar experiments have also 
been performed on the RNA virus $\phi_{6}$ 
\cite{Burch99}. This study tracked the fitness recovery in a population, 
after a deleterious mutation has been induced by a population
bottleneck, for about hundred 
generations. The fitness was seen to recover in steps but the number of steps 
(and the fitness benefit) was found to depend strongly on the population 
size. While large populations recovered in one large step, smaller populations 
required many steps each granting small favors. 
As discussed in detail in the chapter by Ester L\'azaro,
such population bottlenecks occur naturally in the life cycle of viruses, 
because the number of viral particles that are transmitted from one host
to another is often very small.

Finally, we note that
under certain conditions populations of RNA viruses display a linear increase
or decrease of fitness with time \cite{Rouzine03,Novella95}, which can be 
analyzed within the framework of the 
fitness space models discussed in Sect.~\ref{Smooth}.

\section{Conclusions}
\l{Outlook}

In this chapter we have given an overview over a class of models of adaptive 
evolution
which include selection and mutation, but (due to their deterministic 
character)
ignore effects of genetic drift in finite populations. A large body of work 
spread
out over different scientific communities has been devoted to such
models, and our survey must necessarily remain quite incomplete. We have 
therefore
tried to focus on some general concepts -- such as sequence space,
fitness landscapes, error thresholds and epistatic interactions --   
that we believe to be useful 
also beyond the specific biological situations in which the 
models apply.    
 
Stochastic effects characteristic of
finite populations are expected to be quantitatively and even
qualitatively important for several of the phenomena we have described. 
Genetic drift
induces a new mechanism of genetic degradation, \textit{Muller's ratchet} 
\cite{Haigh78}, in which
the fittest genotype is lost from the population because it is not sampled for
reproduction. 
In the limit of infinite sequence length this process is irreversible,
and it generally contributes to the delocalisation of the population from 
fitness peaks. Correspondingly, a common result of finite population studies 
in simple \cite{Wiehe95,Nowak89,Woodcock96} as well as complex 
\cite{Bonhoeffer93,Wilke02b} landscapes is a
lowering of the error threshold mutation rate with decreasing 
population size. A comparison between Muller's ratchet and the
error threshold in infinite population models can be found in 
\cite{Baake00,Wagner93}. 
As described in the chapter by E. L\'azaro,
both mechanisms for genetic degradation are being
considered as possible strategies for fighting viral infections.

The finite size of the population is also crucially important for the peak 
shifts
in rugged landscapes discussed in Sect.~\ref{punctuation}, because it imposes a
cutoff on the tails of rare mutants which are responsible for the communication
between distant fitness peaks. Much of the analytic work on adaptive dynamics 
that takes stochastic
aspects into account has considered the regime of low mutation 
rates\footnote{The quantitative 
characterization of this regime
is that the product of the population size and the mutation probability 
per site is small compared to unity \cite{Wahl00}.}, where the 
population consists of a single genotype at most times and the generation and
fixation of new mutations are rare events. In these studies the geometrical 
constraints
on the availability of new mutants in sequence space are usually ignored, and
the timing and fitness effects of mutations are instead generated by a suitable
stochastic process \cite{Orr00,Gerrish01}. An important task for the future 
will be to integrate
the different theoretical approaches, with the ultimate goal of bringing them
to bear on the experimental data that are becoming available. 

\section*{Acknowledgements}

This work was supported by DFG within 
SFB-TR12 \textit{Symmetries and universality in mesoscopic systems}.
JK is grateful to E. Ben-Naim, D. Krakauer, H. Levine and T. Wiehe
for useful discussions, and to the 
Laboratory of Physics of HUT for the kind hospitality during the completion 
of the article. 

%


%


\printindex
\end{document}